%% file: saem.tex
\documentclass[12pt, a4paper, onecolumn, draftclsnofoot]{IEEEtran}
\usepackage{datetime}
\usepackage{ifpdf}
\usepackage{subfigure}
\usepackage{authblk}
\input{sl.tex}

\ifpdf
  \usepackage[pdftex]{graphicx,color}
  \usepackage[pdftex, colorlinks=true, pdfstartview=FitV,
  linkcolor=blue, citecolor=blue, urlcolor=blue]{hyperref}
\else
  \usepackage{graphicx,color}
  \usepackage[hypertex, colorlinks=true, pdfstartview=FitV, linkcolor=blue,
  citecolor=blue, urlcolor=blue]{hyperref}
\fi
\usepackage{grf}

\usepackage{ulem}
\usepackage{amsmath,amsthm,amscd,amssymb,amsbsy}
\usepackage[mathscr]{eucal}
\usepackage{subfigure,vmargin}

\usepackage{color}

\setpapersize{A4}
\setmarginsrb{1in}{1in}{1in}{1in}{5mm}{5mm}{5mm}{5mm}

\newcommand{\iter}[2]{{#1}^{#2}}

\theoremstyle{plain}
\newtheorem{zdef}{Definition}[section]

\newtheorem{zalg}[zdef]{Algorithm}
\newtheorem{zthm}[zdef]{Theorem}
\newtheorem{zprop}[zdef]{Proposition}
\newtheorem{zlem}[zdef]{Lemma}
\newtheorem{zcoro}[zdef]{Corollary}

\title{String-Averaging Expectation-Maximization for Maximum Likelihood Estimation in Emission Tomography}

\author[1]{Elias Salom\~ao Helou}
\author[2]{Yair Censor}
\author[3]{Tai-Been Chen}
\author[4,5]{I-Liang Chern}
\author[6]{\'Alvaro Rodolfo De Pierro}
\author[7]{Ming Jiang}
\author[8]{Henry Horng-Shing Lu}

\affil[${\tiny 1}$]{\small Department of Applied Mathematics and Statistics, State University of S\~ao Paulo, Postal Box 668, S\~ao Carlos, SP, Brazil. elias@icmc.usp.br.}
\affil[2]{Department of Mathematics, University of Haifa, Mt. Carmel, Haifa 3190501, Israel. yair@math.haifa.ac.il.}
\affil[3]{Department of Medical Imaging and Radiological Sciences, I-Shou University, Kaohsiung City, Taiwan 82445, ROC. ctb@isu.edu.tw.}
\affil[4]{Department of Applied Mathematics, Center of Mathematical Modeling and Scientific Computing, National Chiao Tung University, Hsin Chu, Taiwan 30010, ROC. chern@math.ntu.edu.tw.}
\affil[5]{Department of Mathematics, National Taiwan University, Taipei, Taiwan 10617, ROC. chern@math.ntu.edu.tw.}
\affil[6]{Department of Applied Mathematics and Statistics, State University of S\~ao Paulo, Postal Box 668, S\~ao Carlos, SP, Brazil. alvaro@ime.unicamp.br.}
\affil[7]{LMAM, School of Mathematical Sciences, Beijing International Center for Mathematical Research, Peking University, Beijing 100871, China. ming-jiang@ieee.org.}
\affil[8]{Institute of Statistics, National Chiao Tung University, 1001 University Road, Hsinchu, Taiwan 30010, ROC. hslu@stat.nctu.edu.tw.}

\begin{document}
\maketitle
\normalsize
\begin{center}{May 4, 2013. Revised: December 2, 2013. Revised: January 30, 2014.}\end{center}
\begin{abstract}
  We study the maximum likelihood model in emission
  tomography and propose a new family of algorithms for its
  solution, called String-Averaging Expectation-Maximization
  (SAEM). In the String-Averaging algorithmic regime, the
  index set of all underlying equations is split into
  subsets, called ``strings,'' and the algorithm
  separately proceeds along each string, possibly in
  parallel. Then, the end-points of all strings are
  averaged to form the next iterate. SAEM algorithms with
  several strings presents better practical merits than the classical
  Row-Action Maximum-Likelihood Algorithm (RAMLA). We present
  numerical experiments showing the effectiveness
  of the algorithmic scheme in realistic situations. Performance is
  evaluated from the computational cost and reconstruction quality
  viewpoints. A complete convergence theory is also provided.
\end{abstract}

\newpage

\section{Introduction}
\label{sec:introduction}

The Expectation-Maximization (EM) algorithm (see, e.g.,
\cite[Chapter~7]{lange-book} or \cite{mclahan-book}) has become a
household tool for Maximum Likelihood Estimation (MLE) in
Emission Tomography (ET).
The original EM
algorithm is \textit{simultaneous} since
when passing from a current iterate $x^{k}$ to the next one
$x^{k+1},$ \textit{all} equations $\left\langle a^{i},x\right\rangle
  = b_{i},$ $i=1,2,\cdots,m,$ in the linear system $Ax=b$
of the underlying problem are used. In contrast to this, there is
the Row-Action Maximum Likelihood Algorithm (RAMLA) (of Browne and De
Pierro \cite{browne_pierro_1996}) which is structurally
\textit{sequential} since each iteration $k$ involves
\textit{only one} equation $\left\langle a^{i(k)},x\right\rangle =
b_{i(k)}$, where $i( k )$ is one of the indices $\{ 1, 2, \dots,
m\}$ chosen at the $k$-th iteration,
from the linear system of the underlying
problem. In-between these two structural extremes we have also a
block-RAMLA structure \cite[Page~689]{browne_pierro_1996} and the
Ordered Subsets EM (OSEM) algorithm (of Hudson and Larkin
\cite{hudson_larkin_1994}), both allowing to process in each iteration
a ``block'' (i.e., a subset) of the $m$ underlying equations,
by applying to the equations of the block the simultaneous
original EM iterative step. Section~\ref{sec:EM:type:algorithms} brings a
deeper discussion on alternative methods and related works.

In this article we propose and experiment with a new variant of the EM algorithm
which uses \textit{String-Averaging }(SA)\textit{, }thus we call the resulting
algorithm the \textit{SAEM algorithm}. In the SA algorithmic regime the index
set of the $m$ underlying equations is again split into subsets, now called
``strings,'' and from a current iterate $x^{k}$ the algorithm first proceeds
sequentially along the indices of each string separately (which can be done
in parallel) and then the endpoints of all strings are averaged to form the
next iterate $x^{k+1}.$ This is in stark contrast with how the above mentioned
OSEM and block-RAMLA treat the equations of a block. Full details
regarding the algorithm will be given on Section~\ref{sect:SA}.

We advocate SAEM as a theoretically sound way to provide better quality in
emission tomography image reconstruction. In order to support our claims,
simulation studies performed with the intention to evaluate image quality and
computational effort are reported on Section~\ref{sect:experiments} and a theoretical study
of the convergence properties of the method can be found on Section~\ref{sec:theory}. Next
section introduces the fundamentals of the problem and Section~\ref{sec:conclusions} brings our concluding remarks.

\section{ET image reconstruction: the problem, approaches and related works
\label{sec:related}}

\textit{Positron Emission Tomography} (PET) and \textit{Single
Photon Emission Tomography} (SPECT) are
ideal modalities for in vivo functional
imaging. Thus, ET can be used to
monitor the effects of therapy inside the living body or in oncological studies.
The image reconstruction problem in ET can be fully-discretized and modeled by a
system of linear equations \cite{herman_1980}:
\begin{equation}
Ax=b\label{eq:image:model}%
\end{equation}
where the measured data is $b=(b_{i})_{i=1}^{m}\in\boldsymbol{R}^{m}$,
the image vector is $x=(x_{j})_{j=1}^{n}\in\boldsymbol{R}^{n}$, and $A$ is an $m\times n$
real matrix, and the problem is to estimate the image vector $x$ from the
measured data $b$. Solution of this system by classical direct elimination methods will
likely be impractical because of the huge data and image dimensions, ill-posedness
of $A$ coupled with noisy and/or incomplete data $b$, unstructured imaging matrix $A$,
among several other well-known reasons~\cite{herman_1980, censor_zenios_1997}.

Instead, iterative methods, especially the row-action methods such as the algebraic
reconstruction technique (ART) \cite{kaczmarz_1937, gordon_bender_herman_1970}, are
usually applied for computationally efficient high-quality reconstruction
\cite{herman_meyer_1993}. Iterative algorithms have been important in this field because of
their superior reconstructions over analytical methods in many instances, and their ability
to handle the very large and sparse data associated with fully-discretized image reconstruction,
see, e.g.,~\cite{censor_zenios_1997, fessler_2000,leahy_byrne_2001,jiang_ge_jxst_it_review_2002}.
For reviews see, e.g., \cite{bb96} and \cite{cccdh}.

Another key advantage of iterative methods is their flexibility. For example, least-squares and
{\it I}-divergence optimization approaches are commonly used for image reconstruction.
In these approaches, the true image is estimated by the minimizer of an objective function, i.e.,
as a solution of a convex optimization problem.  The least-squares functional is given by
\begin{equation}
  \label{eq:least:squares}
  L_{\text{LS}} (x) := \frac{1}{2} \| b - A x \|^2,
  \quad\forall x \in\boldsymbol{R}^{n}.
\end{equation}
ART is a prominent algorithm in this approach, which can
be turned to be efficient for ET \cite{herman_meyer_1993} even
though it was originally proposed for x-ray tomography
\cite{kaczmarz_1937, gordon_bender_herman_1970}.  The {\it I}-divergence
(also called the Kullback-Leibler distance) functional is given by
\begin{equation}
L_{\text{KL}}(x):=I(b,Ax)=\sum_{i=1}^{m}b_{i}\log\frac{b_{i}}{{\left\langle
a^{i},x\right\rangle }}+{\left\langle a^{i},x\right\rangle }-b_{i}%
,\quad\forall x\in\boldsymbol{R}_+^{n},\label{eq:kullback:leibler}%
\end{equation}
where the vector $a^{i}=\left( a_{j}^{i}\right) _{j=1}^{n}$ is the
transpose of the $i$-th row of $A$.  From a statistical perspective,
the least-squares approach is equivalent to finding an image
as the maximum likelihood estimate from data with Gaussian noise,
while the {\it I}-divergence approach does the same with Poissonian noise.

For the image reconstruction problem in ET, the {\it I}-divergence approach is
more appropriate because Poissonian noise is dominant. The following iterative
Maximum Likelihood Expectation Maximization recursion,
termed the MLEM (or the classical EM) algorithm,
is popular since the 1980s~\cite{Vardi85} for the minimization
of~\eqref{eq:kullback:leibler}:
\begin{equation}
x_{j}^{k+1}:=x_{j}^{k}\frac{1}{\sum_{i=1}^{m}a_{j}^{i}}\sum_{i=1}^{m}a_{j}%
^{i}\frac{b_{i}}{{\left\langle a^{i},x^{k}\right\rangle }},\quad\text{for
$j=1,2,\cdots,n$}.\label{eq:calssical:em}%
\end{equation}
It was first proposed independently by Richardson~\cite{richardson_1972} and Lucy~\cite{lucy_1974}
and for that reason is still known as the Richardson-Lucy (RL) algorithm in applications to
astronomy and microscopy. It was later rediscovered in \cite{shepp_vardi_1982} and in
\cite{Lange84} by applying the general expectation-maximization (EM) framework from
statistics~\cite{dempster_laird_rubin_1977,bertero_et_al_2009} to the ET reconstruction problem.

In addition to the above two approaches, there are those which
utilize other Bregman divergences (distances) and the Bayesian framework for
image reconstruction \cite{censor_zenios_1997, herman_1980}, of which the
above least-squares and {\it I}-divergence approaches are special cases. In fact, SAEM can
be readily generalized to cover these cases, because our analysis of convergence applies to
the minimization of a general convex function over the nonnegative orthant.

From yet another perspective, the image reconstruction task can also be
viewed as a \textit{convex feasibility problem} (CFP) see \cite{bb96}. Consequently the
method based on projection onto convex sets (POCS) provides significant inspiration when
developing iterative reconstruction algorithms, see, e.g., \cite[Chapter
5]{censor_zenios_1997} or \cite{stark}. There have been various iterative
projection algorithms based on different kinds of projections such as the
Euclidean, oblique and Bregman projections. Among them, the string-averaging
projections (SAP) algorithmic scheme has attracted attention recently since
its presentation in \cite{censor_elfving_herman_2001}. Therefore, the present
paper brings an algorithmic bridge for this string-averaging scheme, going
from the set theoretic to the maximum likelihood framework.

\section{EM-type algorithms\label{sec:EM:type:algorithms}}
\subsection{The classical EM algorithm\label{subsec:EM:classical:EM}}

For $A$, $a^i$, $b$, and $b_i$, as defined above for $i \in
I := \{ 1,2,\cdots,m \}$, we define $R_{\text{EM}}^{i}:$ $\boldsymbol{R}%
^{n}\rightarrow\boldsymbol{R}^{n}$ as the $i$-th row operator for the EM
algorithm,%
\begin{equation}
R_{\text{EM}}^{i}(x):=\frac{b_{i}}{{\left\langle a^{i},x\right\rangle }%
}x.\label{eq:calssical:em:row:op}%
\end{equation}
For any index subset $B\subseteq I$, let $|B|$ denote the number
of elements in $B$ and let $R_{B}$ be the averaging operator $R_{B}:\boldsymbol{R}%
^{|B|\times n}$ $\rightarrow$ $\boldsymbol{R}^{n}$ defined componentwise as%
\begin{equation}
\left(  R_{B}\left(  \{y^{i}\}_{i\in B}\right)  \right)  _{j}:=\frac{1}%
{\sum_{i=1}^{m}a_{j}^{i}}\sum_{i\in B}a_{j}^{i}y_{j}^{i},\quad\text{for
$j=1,2,\cdots,n$}.
\end{equation}

The classical EM algorithm and many of its variants for image reconstruction
are simultaneous algorithms, see, e.g., \cite[Section~1.3]{censor_zenios_1997}
on the classification of iterative algorithms from the perspective of the algorithmic
parallelization strategy. In pseudo-code, the classical EM algorithm, that uses
(\ref{eq:calssical:em}), is as follows:

\begin{zalg}
The Classical EM Algorithm
\begin{align}
&  \text{for $k=0,1,\cdots$}\nonumber\\
&  \qquad\text{for $i\in I$}\nonumber\\
&  \qquad\qquad x^{k+1,i}:=R_{\text{EM}}^{i}(x^{k}%
),\label{eq:calssical:em:row:it}\\
&  \qquad\text{end}\nonumber\\
&  \qquad x^{k+1}:=R_{I}\left(  \{x^{k+1,i}\}_{i\in I}\right)
.\label{eq:calssical:em:it}\\
&  \text{end}\nonumber
\end{align}

\end{zalg}

It first executes simultaneously (in parallel) row-action operations on all
rows from the same current iterate $x^{k}$. Then the intermediate iteration
vectors $\{x^{k+1,i}\}_{i=1}^{m}$ are combined by the averaging operator
$R_{I}$ to obtain the next iterate.

\subsection{The Block iterative EM algorithm\label{subsec:EM:block:EM}}
The classical EM algorithm can be accelerated by its block-iterative version
\cite{hudson_larkin_1994, HQ_Zhu_2004, C_Byrne_2005}. A block-iterative
version of a simultaneous algorithm, see \cite[Section~1.3]%
{censor_zenios_1997}, employs the algorithmic operators $R_{EM}^{i}$ and
$R_{B}$ as the simultaneous algorithm does, but it first breaks up the index set
$I=\{1,2,\cdots,m\}$ into \textquotedblleft blocks\textquotedblright\ of
indices so that for $t=1,2,\cdots,T$, the \textit{block} $B_{t}$ is a subset of
$I$ of the form
\begin{equation}
B_{t}:=\left\{  i_{1}^{t},i_{2}^{t},\cdots,i_{m(t)}^{t}\right\}
.\label{eq:block}%
\end{equation}
with $m(t)$ denoting the number of elements in $B_{t}$. 
The block-iterative version of the classical EM algorithm works as follows.

\begin{zalg}
The Block-Iterative EM Algorithm
\begin{align}
&  \text{for $k=0,1,\cdots$}\nonumber\\
&  \qquad t=t(k)\in\{1,2,\cdots,T\}\\
&  \qquad\text{for $i\in B_{t}$}\nonumber\\
&  \qquad\qquad x^{k+1,i}:=R_{\text{EM}}^{i}(x^{k}),\label{eq:os:em:row:it}\\
&  \qquad\text{end}\nonumber\\
&  \qquad x^{k+1}:=R_{B_{t}}\left(  \{x^{k+1,i}\}_{i\in B_{t}}\right)
.\label{eq:os:em:block:it}\\
&  \text{end}\nonumber
\end{align}

\end{zalg}

At the $k$-th iteration, the active block $B_{t}$ is determined by the
\textit{control sequence} $t=t(k)$.
Then, for each block the algorithm performs a
simultaneous step as if the classical EM method was applied to
this block alone and the iteration is updated.
In the literature of maximum likelihood reconstruction for emission
tomography, block-iterative EM algorithms are called ``ordered-subsets EM''
(OSEM) algorithms as they were named in the seminal work of
Hudson~and~Larkin~\cite{hudson_larkin_1994}.

\subsection{EM-type algorithms with relaxation}

The OSEM algorithm has an improved experimental convergence rate, but
may not converge when the system (\ref{eq:image:model}) is
inconsistent \cite{browne_pierro_1996}. This can be resolved by introducing
relaxation as in the least-squares approach \cite{jiang_ge_sart_os_2003}. The
first block-iterative EM algorithm with relaxation is the \textquotedblleft
row-action maximum likelihood algorithm\textquotedblright\ (RAMLA) in
\cite{browne_pierro_1996}. After statistical study of the noise propagation
from the projection data to the reconstructed image, a modified version of the
RAMLA, called \textquotedblleft dynamic RAMLA\textquotedblright\ (DRAMA),
using variable relaxation parameters with blocks, was proposed in
\cite{tanaka_kudo_2003}. A formula extending RAMLA and DRAMA was proposed in
\cite{neto_depierro_2005} as follows,%
\begin{equation}
x^{k+1}=x^{k}-\lambda_{k,t}D(x^{k})\nabla L_{\text{KL}}^{B_{t}}(x^{k}%
),\label{eq:relaxation:B:os:calssical:em:grad}%
\end{equation}
where $\lambda_{k,t}$ are stepsize parameters, $\nabla L_{\text{KL}}^{B_{t}}$
is the gradient of the partial negative log-likelihood
\begin{equation}
   L_{\text{KL}}^{B_{t}}( x ) := \sum_{i \in B_{t}}b_{i}\log\frac{b_{i}}{{\left\langle
   a^{i}, x \right\rangle}} + {\left\langle a^{i}, x\right\rangle } - b_{i},
\end{equation}
and the pre-conditioner $D$ is given by%
\begin{equation}
D(x):=\operatorname*{diag}\left\{  \left(  \frac{x_{j}}{p_{j}}\right)
\mid1\leq j\leq n\right\}  ,\label{eq:D:RAMLA:DRAMA}%
\end{equation}
with positive constants $p_{j}>0$ . Possible choices of $p_{j}$ are
\begin{align}
  \label{eq:p:j:scaled:grad:descent:maximum}
  p_j &{}= \max_{1 \le t \le T} \left\{ \sum_{i \in B_t} a^i_j \right\},\\
  \intertext{or}
  \label{eq:p:j:scaled:grad:descent:average}
  p_j &{}= \frac{\sum_{i = 1}^m a^i_j}{T}.
\end{align}
Convergence results established in \cite{neto_depierro_2005} require that the following
conditions are met:
\begin{description}[\IEEEsetlabelwidth{Condition A.}]
\item[Condition A:] \label{page:Condition:A} $x^{0}\in\boldsymbol{R}_{+}^{n}$,
$b\in\boldsymbol{R}_{+}^{m}$, $A\in\boldsymbol{R}_{+}^{m\times n}$, $\sum_{t=1}%
^{T}L_{\text{KL}}^{B_{t}}(x)=L_{\text{KL}}(x)$.

\item[Condition B:] \label{page:Condition:B} $\text{rank}\,\left(
W(x)^{\frac{1}{2}}A\right)  =n$, where $W(x)^{\frac{1}{2}}$ is the
component-wise square root of the diagonal matrix%
\begin{equation}
W(x)=\operatorname*{diag}\left\{  \left(  \frac{b_{i}}{{\left\langle
a^{i},x\right\rangle }^{2}}\right)  \mid1\leq j\leq n\right\}  .
\end{equation}
This is equivalent to the strict convexity of the function $L_{\text{KL}}(x)$.

\item[Condition C:] \label{page:Condition:C} $0<\lambda_{k,t}\leq\lambda$,
where $\lambda$ is chosen such that
\begin{equation}
\lambda<\min\left\{  \frac{p_{j}}{\sum_{i\in B_{t}}a_{j}^{i}}\mid
1 \leq j \leq n, 1\leq t \leq T\right\},\label{eq:lambda:p:j}
\end{equation}
and, denoting $\lambda_{k}=\lambda_{k,1}$, such that
\begin{equation}
 \sum_{k=0}^{\infty}\lambda_{k} =\infty;\text{ }\sum_{k=0}^{\infty}\lambda_{k}^{2}
<\infty;\text{ }\sum_{k=0}^{\infty}\left\vert \lambda_{k}-\lambda_{k,t}\right\vert
<\infty;
\label{eq:SUM:Cond:A1}
\end{equation}
\begin{equation}
 \frac{\lambda_{k,t}}{\lambda_{k}}\rightarrow 1; \text{ and }\frac{\lambda_{k}}{\lambda_{k+1}} <\text{constant}.
\label{eq:SUM:Cond:A2}
\end{equation}
\end{description}

The above assumptions on the relaxation parameters are very general and
cover several situations of interest~\cite{tanaka_kudo_2003,ahn_fessler_2003,
ge_jiang_jxst_ossart_2004,tanaka_kudo_2010}. If each block size is one,
i.e., each block corresponds to one equation, then by using
\eqref{eq:relaxation:B:os:calssical:em:grad} and
\eqref{eq:D:RAMLA:DRAMA}, we get the relaxed row-action
iteration of RAMLA:
\begin{equation}
x_{j}^{k+1}=x_{j}^{k}+\lambda_{k,i}\frac{a_{j}^{i}}{p_{j}}\left(  \frac{b_{i}%
}{{\left\langle a^{i},x^{k}\right\rangle }}-1\right)  x_{j}^{k},\text{ for
$j=1,2,\cdots,n$}.\label{eq:relaxation:row:action:em}%
\end{equation}

\section{The string-averaging scheme\label{sect:SA}}
\subsection{The string-averaging prototypical scheme}
The \textit{string-averaging} (SA) algorithmic regime was originally
formulated in \cite{censor_elfving_herman_2001} in general terms and applied
there for solving the \textit{convex feasibility problem} (CFP) with iterative
projection algorithms, see, e.g., \cite{censor_elfving_herman_2001,
censor_tom_2003, penfold_et_al_2010}. In the string-averaging paradigm, the
index set $I=\{1,2,\cdots,m\}$ is split into \textquotedblleft
strings\textquotedblright\ of indices. From the current iterate $x^{k}$,
certain algorithmic operators (we shall call them \textit{step operators} in
the following) are applied sequentially along the indices of each string and
the end-points of all strings are then combined by an additional algorithmic
operator (which we name the \textit{combination operator} from now on)
to yield the next iteration vector.

To define SA algorithms precisely, the same decomposition as in
(\ref{eq:block}) for the index set $I$ is utilized. The \textit{index set}
\begin{equation}
  \label{eq:S:t}
  S_{t}
    :=\left\{ i_{1}^{t},i_{2}^{t},\cdots,i_{m(t)}^{t}\right\}
    = \left\{ i_{s}^{t}|\ 1 \le s \le m(t)\right\}
\end{equation}
is now serving as the \textit{string} of indices in the
current context, for $t=1,2,\cdots,T$. Viewed like that,
strings and blocks are just names for index subsets
$B_{t}\subseteq I$ or $S_{t}\subseteq I$ for
$t=1,2,\cdots,T$.
Interleaving of strings and blocks is possible, leading to
algorithms with a tree-like parallelism structure,
whose convergence properties can be almost directly devised
from our analysis below.

Let us consider a set $Q\subset\boldsymbol{R}^{n}$ and
family of operators $\{R^{i}\}_{i=1}^{m}$ mapping $Q$ into
itself, and an additional operator $R$ which maps $Q^{T}$
(i.e., the product of $T$ copies of $Q$) into $Q$. In the
SA paradigm these operators $\{R^{i}\}_{i=1}^{m}$ are
the step operators and the operator $R$ serves as
the combination operator. The string-averaging
prototypical scheme is as follows.

\begin{zalg}
\label{string:averaging:prototypical:algorithmic:scheme}
The String-Averaging Prototypical Scheme
\cite{censor_elfving_herman_2001}

\begin{description}[\IEEEsetlabelwidth{Iterative Step}]
\item[Initialization] $x^{0} \in Q$ is an arbitrary starting point.

\item[Iterative Step] Given the current iterate $x^{k}$,

\begin{description}[\IEEEsetlabelwidth{(ii)}]
\item[(i)] for all $t=1,2,\cdots,T$, compute in parallel as
    follows: apply successively the step operator along the
    string $S_t$,
  \begin{equation}
    x^{k+1,t}:=R^{i_{m(t)}^{t}}\circ\cdots\circ R^{i_{2}^{t}}\circ R^{i_{1}^{t}%
    }(x^{k}),\label{string:av:1}%
  \end{equation}

\item[(ii)] apply the combination operator
  \begin{equation}
    x^{k+1}:=R\left(  \{x^{k+1,i}\}_{t=1}^{T}\right)  .\label{string:av:2}%
  \end{equation}
\end{description}
\end{description}
\end{zalg}

For every $t = 1, 2, ..., T$, this algorithmic scheme first
applies to $x^{k}$ successively the step operators
$R^{i}$ whose indices $i$ belong to the $t$-th string
$S_t$. This can be done in parallel for all strings
because the jobs of going along each string from (one and
the same current iterate) $x^{k}$ to the end-point
$x^{k+1, t}$ are independent. Then the combination
operator $R$ maps all end-points onto the next
iterate $x^{k+1}$. The iteration from $x^{k}$ to $x^{k+1}$
is called one cycle of iteration, whereas the iteration
from $x^{k}$ to $x^{k+1, t}$ is called the
$t$-th sub-iteration in the $k$-th cycle. Notice that we
can always obtain from this framework a
\textit{fully-sequential} algorithm by the choice $T=1$ and
$S_{1} = I$ or a \textit{fully-simultaneous} algorithm by
the choice $T = m$ and $S_{t}=\{t\}$, $t = 1, 2, \cdots, T$.

\subsection{The string-averaging EM algorithm}
\label{subsec:SA:EM}

Here we merge the string-averaging algorithmic structure
describe above with the maximum likelihood estimator in order
to create the new string-averaging EM (SAEM) algorithm. To
this end we adopt the row-action operation of RAMLA as step
operators (\ref{eq:relaxation:row:action:em}), namely:
\begin{equation}
( R_{\lambda_{ k, i }}^{i}( x^{k} ) )_{j}
:= x_{j}^{k}+\lambda_{k,i}\frac{a_{j}^{i}}{p_{j}}\left(
\frac{b_{i}}{{\left\langle a^{i},x^{k}\right\rangle }}-1\right)  x_{j}%
^{k},\text{ for $j=1,2,\cdots,n$}.\label{eq:ramla-step}%
\end{equation}
To combine end-points of
strings we use convex combinations, thus, our algorithm
takes the following form.

\begin{zalg}
\label{alg:SAEM}The \textit{String-Averaging-EM} (SAEM)

\begin{description}[\IEEEsetlabelwidth{Iterative Step}]
\item[Initialization] Choose parameters $p_{j},$ for $j=1,2,\cdots,n,$ and parameters
$\lambda_{k,i}$. Choose $x^{0}>0$ as an arbitrary initial vector, construct
a family of strings $\{S_{t}\}_{t=1}^{T}$, choose a weight system $\{w_{t}\}_{t=1}%
^{T}$ such that $w_{t}>0$ for all $t=1,2,\cdots,T$, and $\sum_{t=1}^{T}%
w_{t}=1$.

\item[Iterative Step] Given the current iterate $x^{k}$,

\begin{description}[\IEEEsetlabelwidth{(ii)}]
\item[(i)] for all $t=1,2,\cdots,T$, compute (possibly in parallel) as
    follows: apply successively the RAMLA row-action iterative step
    given by (\ref{eq:ramla-step}) along the string $S_t$,
\begin{equation}
x^{k+1,t}
:= R^{S_t}(x)
:=R^{i_{m(t)}^{t}}\circ\cdots\circ R^{i_{2}^{t}}\circ R^{i_{1}^{t}%
}(x^{k}),\label{sa:em:1}%
\end{equation}
where the dependence of each step operator in
(\ref{sa:em:1}) on the relaxation parameter $\lambda_{k,
i^t_s}$, $s = 1, 2, \cdots, m(t)$ has been left out for clarity.

\item[(ii)] then combine the end-points by%
\begin{equation}
x^{k+1}=\sum_{t=1}^{T}w_{t}x^{k+1,t}\label{sa:em:2}%
\end{equation}
using the weights system $\{w_{t}\}_{t=1}^{T}$.
\end{description}
\end{description}
\end{zalg}

\section{Theoretical Justification}\label{sec:theory}

In this section we provide a general justification
for why SAEM algorithms present a convergent behavior. We
base our proofs on the fact that the convex combination operator,
used in the averaging step, preserves certain asymptotic
characteristics of the step operator and, therefore, we can
expect convergence whenever the step operator comes from a
convergent algorithm as is the case here.

Incremental algorithms \cite{heloudep}, such as those used
for the stringing by the step operator of the
SAEM algorithm satisfy an approximation given by
\begin{equation}
  R^{S_t}(x,\lambda)=x-\lambda D(x)\nabla L^{S_t}(x)+O(\lambda^2)\text,
\label{eq:generalstep}
\end{equation}
according to Proposition~\ref{prop:approx} below. In the
current theory, the general form (\ref{eq:generalstep})
allows us to prove convergence under suitable hypotheses for
non-averaged algorithms, mainly because $L^{S_t}=L$ and,
therefore, for such iterations we have:
\begin{equation}
  \label{eq:non:averaged:general:step}
  \iter x{k + 1} = \iter xk - \lambda_k D( \iter xk )
  \nabla L( \iter xk ) + O( \lambda_k^2 ).
\end{equation}
But any conclusion drawn from equation
\eqref{eq:non:averaged:general:step} should hold as well for
a properly averaged algorithm, by replacing $L$ by
the following $\tilde L$:
\begin{equation}
  \label{eq:tilde:L}
  \tilde L (x):=\sum_{t=1}^Tw_tL^{S_t} (x),
\end{equation}
where $w_t > 0$ are the
  weights. For the SAEM algorithm~\ref{alg:SAEM}
with $\sum w_i=1$, we have
\begin{equation}
  \iter x{k+1}
  =\iter xk-\lambda_k D(\iter xk)\nabla\tilde L(\iter xk)+O(\lambda_k^2).
\end{equation}
If the strings $S_t$ are pairwise disjoint and such that
$\bigcup_{t=1}^TS_t=\{1, 2, \dots,m\}$ and $w_t=1/T$, we have
\begin{equation}\label{eq:iterlambdaoverT}
  \iter x{k+1}
  =\iter xk-\frac{\lambda_k}T D(\iter xk)\nabla L(\iter xk)+O(\lambda_k^2)
\end{equation}
for the averaged iteration, and convergence will be towards the
optimizer of $\tilde L = L$. In order to make the
discussion more precise, from now on we consider the
following general form of an averaging algorithm:
\begin{zalg}\label{algo:avgeneral}
	General Averaging Algorithm
	\begin{align}
	&  \text{for $k=0,1,\cdots$}\nonumber\\
	&  \qquad\text{for $t=1,2,\dots,T$}\nonumber\\
	&  \qquad\qquad x^{k+1,t}:=R^{S_t}(x^{k},\lambda_k),\nonumber\\
	&  \qquad\text{end}\nonumber\\
	&  \qquad x^{k+1}:=\sum_{t=1}^T\omega_tx^{k+1,t}\nonumber\\
	&  \text{end}\nonumber
	\end{align}
\end{zalg}
By considering this algorithm we cope with step
operators ${R}^{S_t}$ of a rather general form, but stick
to the concrete realization of the combination operator.
For this kind of iteration we have the following result.
\begin{zprop}\label{prop:propagaprpox}
  Suppose that $\omega_{t}> 0$ and $\sum_{t=1}^{T}\omega_t=1$ in Algorithm~\ref{algo:avgeneral}, and that
for every $t\in\{1, 2, \dots,T\}$ the following equality holds for all $k\geq 0$,
  \begin{equation}\label{eq:hipapprox}
    R^{S_t}(\iter xk,\lambda_k)
    =
    \iter xk
    -
    \lambda_kD(\iter xk)\nabla L^{S_t}(\iter xk)+O\bigl(\Gamma(\lambda_k)\bigr).
  \end{equation}
  Then we have
  \begin{equation}
    \iter x{k+1}
    =
    \iter xk
    -
    \lambda_kD(\iter xk)\nabla\tilde L(\iter xk)+O\bigl(\Gamma(\lambda_k)\bigr),
  \end{equation}
where $\tilde L$ is defined in (\ref{eq:tilde:L}).
\end{zprop}
\begin{IEEEproof}
  The result is verified by observing that
 \begin{align}
    \iter x{k+1}
     &= \sum_{t=1}^{T}\omega_tR^{S_t}(x^{k},\lambda_k)\nonumber\\
     &= \sum_{t=1}^{T}\omega_t\iter xk
        - \lambda_kD(\iter xk) \nabla\sum_{t=1}^{T}\omega_tL^{S_t}(\iter xk)
        +\sum_{t=1}^{T}\omega_tO\bigl(\Gamma(\lambda_k)\bigr)\nonumber\\
     &= \iter xk - \lambda_kD(\iter xk)\nabla\tilde L(\iter xk)
       + O\bigl(\Gamma(\lambda_k)\bigr),
\end{align}
where we have used the definition of the algorithm for the first
  equality, hypothesis~\eqref{eq:hipapprox} for the second, and then
  applied $\sum_{t=1}^{T}\omega_t=1$, the definition of $\tilde L$, and
  a trivial property of the $O$ notation to obtain the third equation.
\end{IEEEproof}

We now prove the claim that our stringing operators
${R}^{S_t}$ do satisfy an equation
such as~\eqref{eq:hipapprox}
with $\Gamma(\lambda_k)=\lambda_k^2$, as long as
Lipschitz continuity holds for each parcel of
$D(\iter xk)L^{S_t}(\iter xk)$ used during each
stringing operation, according to the next statement.

\begin{zprop}\label{prop:approx}
  Let $L^{S_t}=\sum_{i \in S_t} L^{i}$ and
  $R^{S_t}(x,\lambda):=R^{i_{m(t)}^t}_\lambda\circ\cdots\circ
  R^{i_{2}^t}_\lambda\circ R^{i_{1}^t}_\lambda(x)$, where
each $R^{i}_\lambda$ for $i \in \{1, 2, \cdots,   m \}$ is
given by
\begin{equation}
  \label{eq:general:R:i}
  R^{i}_\lambda(x):=x-\lambda D(x)\nabla L^{i}(x)
\end{equation}
For $k = 1, 2, \cdots$, denote
\begin{align}
  \label{eq:y:k:j:0}
  \iter y{k,0} & := \iter xk, \\
  \iter y{k,s} & := R^{i_{s}^t}_{\lambda_k}(\iter y{k,s-1}), \quad
  s = 1, 2, \cdots, m(t), \\
  x^{k+1,t} & :=  \iter y{k,m(t)}\\
  \iter x{k+1} & := \sum_{t=1}^T\omega_t x^{k+1,t}.
\end{align}
Assume that each $D(\cdot)\nabla L^{i_{j}^t}(\cdot)$ is
Lipschitz continuous with a Lipschitz constant $M$ and bounded by an
upper bound $N$ on the set $\{\iter xk,\iter y{k,s}\}_{k\in\mathbb N,
  s \in I}$, then the operator $R^{S_t}$ satisfies
\begin{equation}
  R^{S_t}(\iter xk,\lambda_k)=\iter xk-\lambda_kD(\iter xk)\nabla L^{S_t}(\iter xk)+O(\lambda_k^2).
\end{equation}
\end{zprop}
\begin{IEEEproof}
  Unfolding of the definition of the operator leads to
  \begin{equation}\label{eq:defioperaux}
    R^{S_t}(\iter xk,\lambda_k)=\iter xk-\lambda_k\sum_{s=1}^{m(t)}D(\iter y{k,s-1})\nabla L^{i^t_s}(\iter y{k,s-1}).
  \end{equation}
  Consider the magnitude of the following difference,
  \begin{equation}\label{eq:defidifaux}
    \iter\epsilon k
    :=
    D(\iter xk)\nabla L^{S_t}(\iter xk)
    -
    \lambda_k\sum_{s=1}^{m(t)}
    D(\iter y{k,s-1})\nabla L^{i^t_s}(\iter y{k,s-1}),
  \end{equation}
  which we can estimate as follows,
  \begin{align}
    \left\|\iter\epsilon k\right\|
    &=
    \left\|
    \sum_{s=1}^{m(t)}\left( D( \iter xk )
\nabla L^{i^t_s}( \iter xk ) -
      D(\iter y{k,j-1})\nabla L^{i^t_s}( \iter y{k,j-1} )
      \right)\right\|\\
         &\leq\sum_{s=1}^{m(t)}
    \left\|
      D(\iter xk)\nabla L^{i^t_s}(\iter xk)
      -
      D(\iter y{k,j-1})\nabla L^{i^t_s}(\iter y{k,j-1})
    \right\|\\
    &\leq\sum_{s=1}^{m(t)}M\left\|\iter xk-\iter y{k,s-1}\right\|,\label{eq:dif1}
  \end{align}
  where the first inequality follows from the triangular inequality,
  and the latter from Lipschitz continuity. Now
  we notice that if $s > 1$, then
  \begin{equation}\label{eq:dif2}
    \|\iter xk-\iter y{k,s}\|
    =
    \left\|
      \lambda_k
      \sum_{l=1}^{s}
      D(\iter y{k,l-1})\nabla L^{i_t^j}(\iter y{k,l-1})
    \right\|
    \leq
    \lambda_k
    m N
  \end{equation}
  because of the boundedness hypothesis. Putting
  \eqref{eq:dif1}~and~\eqref{eq:dif2} together leads to
  $\|\iter\epsilon k\|\leq\lambda_kK$ for some large enough constant $K$, i.e., $\iter\epsilon k=O(\lambda_k)$. On the other
  hand, from \eqref{eq:defioperaux}~and~\eqref{eq:defidifaux}
  \begin{equation}
    R^{S_t}(\iter xk,\lambda_k)=\iter xk-\lambda_kD(\iter xk)\nabla L^{S_t}(\iter xk)+\lambda_k\iter\epsilon k,
  \end{equation}
  which implies the assertion, since $\lambda O(\lambda)=O(\lambda^2)$.
\end{IEEEproof}

We conclude by making the following connection.

\begin{zcoro}\label{coro:approxiter}
  Under the assumptions of Propositions~\ref{prop:approx} and
  \ref{prop:propagaprpox}, Algorithm~\ref{algo:avgeneral}
  satisfies, for all $k\geq0$,
  \begin{equation}\label{eq:approxiter}
    \iter x{k+1}=\iter xk-\lambda_kD(\iter xk)\nabla\tilde L(\iter xk)+O(\lambda_k^2).
  \end{equation}
\end{zcoro}
\begin{IEEEproof}
  Use Proposition~\ref{prop:approx} followed by Proposition \ref{prop:propagaprpox}.
\end{IEEEproof}

So far we have shown how the string averaged
iteration approximates a scaled gradient iteration. From
now on we study how to obtain convergence from
these approximate steps, which requires conditions on
the sequence of parameters
$\{\lambda_k\}\subset\boldsymbol R_+$. We will
need $\sum_{k=1}^\infty\lambda_k=\infty$ because we
will usually assume some sort of boundedness on
$D(\iter xk)\nabla L(\iter xk)$ and, consequently, we
would never reach optimal points if they happen to be too
far away from the initial guess. On the other hand, if we
do not impose $\lambda_k\to0$, the error $O(\lambda_k^2)$
in the approximation does not necessarily vanish and
will eventually dominate the computations when
$\|D(\iter xk)\nabla L(\iter xk)\|$ becomes small. This
must happen if we wish to converge to an optimal point
since a necessary condition on a solution $x^*$ for
the non-negatively constrained problem is $D(x^*)\nabla
L(x^*)=0$. We now show that, with no further hypotheses, at
least one subsequence generated by the algorithm is of
interest.

We will need the following lemma to prove the next
proposition.
\begin{zlem}
  \label{lem:1}
  (Lemma~2, \cite{trummer_1981})
  Assume that $\{ e_k \}$, $\{ \nu_k \}$ and $\{ d_k \}$ are sequences of
  non-negative numbers satisfying for all $k \geq 0$,

\begin{equation}
  e_{k+1} \le e_k - \nu_k d_k
\end{equation}
and that
  $\sum_{k=0}^{\infty} \nu_k = + \infty$.
  Then $0$ is a cluster point of $\{ d_k \}$.
\end{zlem}

\begin{zprop}
  \label{prop:convweak}
  Suppose that the algorithm generating $\{\iter xk\}_{k\in\mathbb N}$
  satisfies~\eqref{eq:approxiter}, $\sum_{k\in\mathbb
    N}\lambda_k=\infty$, $\lambda_k\to0^+$, $\tilde L$ is smoothly
  differentiable and $\tilde L(\iter xk)$ is well-defined.
  Assume also that $\{\iter xk\}\subset\boldsymbol R_+^n$ is
    bounded.  Then either $\lim\tilde L(\iter
  xk)=-\infty$ (i.e., the sequence of objective values is unbounded
  from below) or there is a subsequence $l_k$ such that
  \begin{equation}
    \label{eq:D:nabla:L=0}
    \lim_{l_k\to\infty}
    D(x^{l_k}) \nabla \tilde L(x^{l_k})=0.
  \end{equation}
\end{zprop}

\begin{IEEEproof}
We first estimate how much does the objective function improve at
each iteration. By using~\eqref{eq:approxiter} followed by a
first-order Taylor expansion we get:
  \begin{align}
    \tilde L(\iter x{k+1})&=\tilde L\left(\iter x{k}-\lambda_kD(\iter xk)\nabla\tilde L(\iter xk)+O(\lambda_k^2)\right)\nonumber\\
    &=\tilde L(\iter x{k})-\lambda_k\nabla\tilde L(\iter xk)^TD(\iter xk)\nabla\tilde L(\iter xk)+O(\lambda_k^2)\text.
  \end{align}
Then we rewrite the above equality as the following inequality, holding for some $M>0$, by
definition of $O( \lambda^2 )$:
\begin{equation}\label{eq:majorlkp1}
\tilde L(\iter x{k+1}) \le \tilde L(\iter x{k})-\lambda_k(\nabla\tilde L(\iter xk)^TD(\iter xk)\nabla\tilde L(\iter xk)
-M\lambda_k)
\end{equation}
Now suppose that $\nabla\tilde L(\iter xk)^TD(\iter xk)\nabla\tilde L(\iter xk)$
is bounded from below by a positive $\beta$. Then, for $k$ large enough,
the factor between parentheses in the second term of~\eqref{eq:majorlkp1} will be
nonnegative, satisfying the nonnegativity hypotheses of Lemma~\ref{lem:1}.
Therefore, if the sequence of objective values is bounded from below,
it follows from Lemma~\ref{lem:1} that there is a subsequence
$l_k$ such that

\begin{equation}
  \label{eq:D:nabla:L=0:1}
  \lim_{k\to\infty}
  \left\{
  \nabla\tilde L(\iter x{l_k})^T D(x^{l_k}) \nabla \tilde L(x^{l_k})
  +
  O(\lambda_{l_k})
  \right\}
  =0.
\end{equation}
Because $\lambda_k\to0^+$, we have
\begin{equation}
  \label{eq:D:nabla:L=0:2}
  \lim_{k\to\infty}
  \nabla\tilde L(\iter x{l_k})^T D(x^{l_k}) \nabla \tilde L(x^{l_k})
  =0.
\end{equation}
Consequently, for each $j = 1, 2, \cdots, n$,
\begin{equation}
  \label{eq:D:nabla:L=0:3}
  \lim_{l_k\to\infty}
  D_j(x^{l_k}) \left(\partial_j \tilde L(x^{l_k}) \right)^2 =0.
\end{equation}
By the boundeness assumption, we can assume that, with the same
subsequence $\{l_k\}$,
\begin{equation}
  \label{eq:D:nabla:L=0:4}
  \lim_{l_k\to\infty}
  D (x^{l_k}) = d,
\end{equation}
for some $d \in \boldsymbol R^{n \times n }$. Hence, we
obtain, for each $j = 1, 2,
\cdots, n$,
\begin{equation}
  \label{eq:D:nabla:L=0:5}
  \lim_{l_k\to\infty}
  \left( D_j(x^{l_k}) \right)^2
  \left(\partial_j \tilde L(x^{l_k}) \right)^2
  =
  d_{jj} \cdot 0 = 0.
\end{equation}
Therefore, for each $j = 1, 2, \cdots, n$,
\begin{equation}
  \label{eq:D:nabla:L=0:6}
  \lim_{k\to\infty}
  D_j(x^{l_k}) \partial_j \tilde L(x^{l_k})
  = 0,
\end{equation}
from which the conclusion follows.
\end{IEEEproof}

In particular, if it is known beforehand that $\tilde L$ is
bounded from below, then we have shown that the algorithm does
provide, at least, a useful subsequence. This is likely to be the case,
since in real optimization problems (like in
tomographic reconstruction), one will not be able to obtain
an unboundedly good solution. Using the above result with
some further assumptions (which include strict convexity)
we obtain the following global convergence result.

\begin{zcoro}\label{coro:conv_str_hyp}
Suppose that the assumptions of Propositions~\ref{prop:approx}
  and \ref{prop:propagaprpox} hold, assume also that $\tilde L$ is
smoothly differentiable and strictly convex and that $\tilde
  L(\iter xk)$ is well-defined and that $\{\iter xk\}\subset\boldsymbol
R_+^n$ is bounded. If $\tilde L(\iter xk)$ converges, then the
sequence generated by Algorithm~\ref{algo:avgeneral} converges to the
solution of
    \begin{equation}
        \min_{x\in\boldsymbol R_+^n}
        \quad\tilde L(x)\nonumber\text.
        \label{eq:probnoneg}
    \end{equation}
\end{zcoro}
\begin{IEEEproof}
  If $\{\iter xk\}$ is bounded then we may assume, without
  loss of generality, that the subsequence which we have proven to exist in
  Proposition~\ref{prop:convweak} is convergent, say $\iter
  x{l_k}\to x^*$. If $x^*\in\boldsymbol R_+^n$ and $\tilde L$ is
  convex, then $D(x^*)\tilde L(x^*)=0$ is a necessary and sufficient
  condition for $x^*$ to be an optimal solution of
  problem~\eqref{eq:probnoneg}.

  By continuity, because $\tilde L(\iter x{l_k})$
  converges, it must converge to $\tilde L(x^*)$.  Therefore, the objective value at every accumulation point of
  $\{\iter xk\}$ equals $\tilde L(x^*)$.  However, since we assumed
  that $\tilde L$ is strictly convex, this reasoning implies that
  all accumulation points of the bounded sequence $\{\iter xk\}$ are
  the same, namely $x^*$. Thus, we have proven that $\iter xk\to
  x^*$.
\end{IEEEproof}

We now remove some of the hypotheses used in the last corollary by referring
to works available in the literature. This leads to our main result,
stated below.

\begin{zthm}
   Assume that all components of $\iter x0\in\boldsymbol R_+^n$ are positive, $0<\lambda_k\leq\lambda$ for
some suitably small $\lambda>0$, $\sum_{k=0}^\infty\lambda_k=\infty$ and
$\sum_{k=0}^\infty\lambda_k^2<\infty$. Then, if $L$ is
the negative Poisson log-likelihood and if $\tilde L$ is strictly convex,
then the sequence generated by Algorithm~\ref{algo:avgeneral} converges
to the solution of~\eqref{eq:probnoneg}.
\end{zthm}
\begin{IEEEproof}
   Boundedness of $\{x^k\}$, for small enough $\lambda_k$, is a consequence of~\cite[Proposition 1]
{browne_pierro_1996}. Convergence of $\tilde L(\iter xk)$ may then be obtained by assuming that
$D(x)\nabla\tilde L(x)$ is Lipschitz continuous on the closure of the convex hull of the iterates and
that $\sum_{k=0}^\infty\lambda_k^2<\infty$, as in ~\cite[Lemma 3]{ahn_fessler_2003}. Furthermore, we see
that, if $\lambda_k\leq\lambda$ for some suitably small $\lambda>0$, it is possible to adapt the ideas
leading to~\cite[Corollary 1]{neto_depierro_2005} to our case, yielding the conclusion that
$D(x)\tilde L(x)$ is Lipschitz continuous within the closure of the convex hull
of the iterates (and subiterates) of the averaged algorithm~\eqref{algo:avgeneral}. Therefore,
since $\tilde L$ is naturally bounded from below, we can apply Corollary~\ref{coro:conv_str_hyp}.
\end{IEEEproof}

\section{Experimental work\label{sect:experiments}}

This section is devoted to the experimental setup we
have used in order to test the practical usefulness of SAEM
algorithms and the results obtained. We will
base our conclusions in two different figures of merit, the
first is the squared error from the ground truth image, and the
second being a Total-Variation based analysis, which puts on firmer
grounds our claims about image quality.  In what follows we provide a
detailed description of the experimental setup.
Subsection~\ref{subsec:ira} will report the experimental
results.

There are certain sources of pseudo-randomness in the
numerical experiments we have performed, such as the choice
of strings and the noise simulations,
but the presented results are representative of the typical
situation, as we have noticed little deviation from this
behavior from one run to another of the experiments.
Moreover, to enable any interested reader to reproduce our
research, the full source code of the numerical algorithms
used in this paper is available upon request.

\paragraph{Data Generation}

\begin{figure}
   \centering%
   \input{figradonteta.tex}%
   \def\tlinha{-0.6}%
   \setlength{\grftotalwidth}{0.45\columnwidth}%
   \setlength{\grfticksize}{0.5\grfticksize}%
   \small{\ }\hfill%
   \begin{grfgraphic}{%
      \def\grfxmin{-2.05}\def\grfxmax{1.5}%
      \def\grfymin{-1.5}\def\grfymax{2.05}%
   }%
      \begin{scope}[>=stealth,style=grfaxisstyle,<->]%
         \shepplogan%
         \draw (-1.5,0) -- (1.5,0);%
         \draw (0,-1.5) -- (0,1.5);%
         \foreach \i in {-1,1}%
         {
            \draw[style=grftickstyle,-] (\i\grfxunit,-\grfticksize) -- (\i\grfxunit,\grfticksize);%
            \draw[style=grftickstyle,-] (-\grfticksize,\i\grfyunit) -- (\grfticksize,\i\grfyunit);%
         }
         \begin{scope}[rotate=\teta]
            \draw (-1.5,0) -- (1.5,0);
            \foreach \i in {-1,1}
               \draw[style=grftickstyle,-] (\i\grfxunit,-\grfticksize) -- (\i\grfxunit,\grfticksize);
            \draw[dashed,dash phase=-0.005\grfyunit] (\tlinha,-1.5) -- (\tlinha,1.5);
            \fill (\tlinha,0) node[anchor=north,inner sep=\grflabelsep] {\scriptsize$t$} circle (0.025cm);
            \draw[-] (\tlinha\grfxunit,0.3em) -| (\tlinha\grfxunit - 0.3em,0pt);
            \fill (\tlinha\grfxunit - 0.15em,0.15em) circle (0.025cm);
         \end{scope}
         \begin{scope}[rotate=\teta,yshift=1.5\grfyunit]
            \draw[line width=0.025cm] plot file {figradon.data};
            \draw[-]  (1.5,0) -- (-1.5,0) node[anchor=north,rotate=\teta,inner sep=\grflabelsep] {\scriptsize$t$};
            \draw[->] (0,0)    -- (0,0.7) node[anchor=west,rotate=\teta,inner sep=\grflabelsep]  {\scriptsize$\mathcal R[f](\theta,t)$};
         \end{scope}
         \def\rad{0.075}
         \FPupn{\cpt}{0.552285 \rad{} * 90 \teta{} / *}
         \path (\teta:\rad) ++(\teta - 90:\cpt) node (a) {};
         \draw[-] (0,0) -- (\rad,0) .. controls +(0,\cpt) and (a) .. (\teta:\rad) -- cycle;
         \draw[style=grftickstyle,-,rotate=\tetameio,xshift=\rad\grfxunit] (-\grfticksize,0pt) -- (\grfticksize,0pt);
         \path[rotate=\tetameio,xshift=0.3em] (\rad,0) node[anchor=west,inner sep=0pt] {\scriptsize$\theta$};
      \end{scope}
   \end{grfgraphic}\hfill%
   \begin{grfgraphic}[1.125]{%
      \grfyaxis[R]{[]-1;0;1[]}{[]\tiny$-1$;\tiny$0$;\tiny$1$[]}%
      \grfylabel{\footnotesize$t$}%
      \grfxaxis[R]{[]0;1.57;3.14[]}{[]\tiny$0$;\tiny$\frac\pi2$;\tiny$\pi$[]}%
      \grfxlabel{\footnotesize$\theta$}%
      \def\grfxmin{0}%
      \def\grfxmax{3.14}%
      \grfwindow%
   }%
      \node[anchor=north west,inner sep=0pt] at (0,1)
      {\includegraphics[width=3.14\grfxunit,height=2\grfyunit]
      {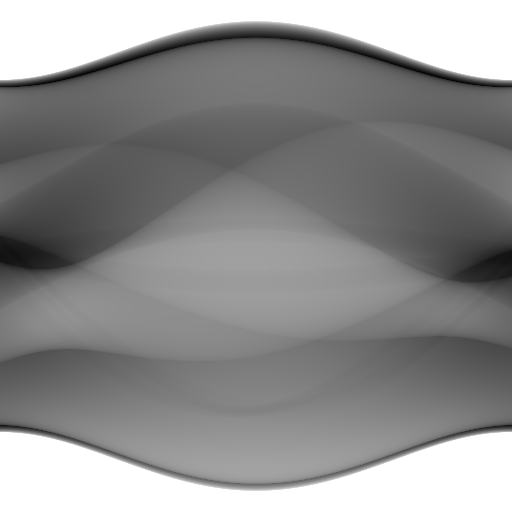}};%
   \end{grfgraphic}\hfill{\ }%
   \caption{Left: schematic representation of the Radon transform. In
   the definition, $\theta$ is the angle between the normal to
   the integration path and the vertical axis, while $t$ is the
   displacement from origin of the line of integration.
   Right: Image of the Radon transform of the image shown on the
   left in the $\theta \times t$ coordinate system.}\label{fig:rad}%
\end{figure}
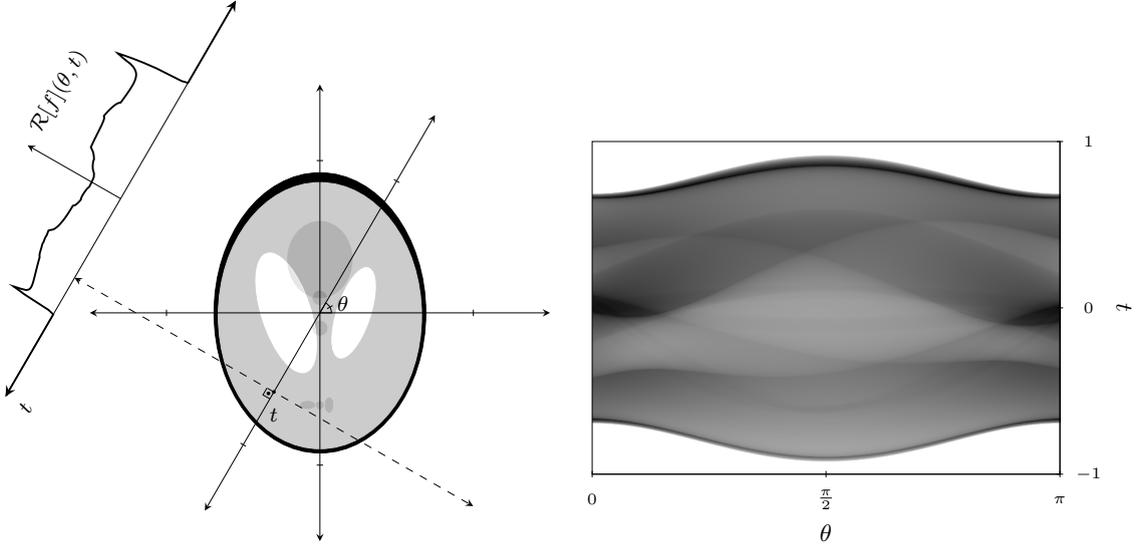

In our simulated studies we used the modified Shepp-Logan head
phantom~\cite{shepp_vardi_1982} in order to compare the
quality of images reconstructed by
RAMLA and SAEM. Let us define the so-called Radon transform
$\mathcal R[f]$ of a function $f: \mathbf R^2 \mapsto \mathbf
R$:
\begin{equation*}
   \mathcal R[ f ]( \theta, t ) := \int_{\mathbf R}
   f\bigl( t( \cos\theta, \sin\theta )^T + s( -\sin\theta,
   \cos\theta )^T \bigr) \mathrm ds.
\end{equation*}
Figure~\ref{fig:rad} shows a schematic representation of the
basic geometry of the Radon transform. The Shepp-Logan phantom
is shown, together with its \emph{sinogram}, i.e., its Radon
transform presented as an image in the $\theta \times t$
coordinate system.

\input{exp1000_sub_data.tex}
We have used
a parallel beam geometry where the samples
were measured at the pairs $( \theta, t )$ in the set
$\{\theta_1, \theta_2, \dots \theta_v\} \times \{t_1, t_2,
\dots, t_r\}$ where $\theta_i = \pi( i - 1 ) / v$ and
$t_j = -1 + 2( j - 1 )/( r - 1 )$ for $1 \leq i \leq v$ and
$1 \leq j \leq r$. That is, we uniformly sample the Radon
transform on the box $[ 0, \pi ) \times [ -1, 1 ]$.
Data was generated as samples of a Poisson
random variable having as parameter the exact Radon
transform of the scaled phantom:
\begin{equation}
   b_i \sim Poisson \bigl( \kappa \mathcal R[ f ](
   \theta_{v_i}, t_{r_i} ) \bigr), \quad 1 \leq i \leq vr\text,
\label{eq:image:model:TB}%
\end{equation}
where $v_i$ is one plus the largest integer smaller than
$(i - 1) / r$, $r_i = i - r(v_i - 1)$, and where the scale factor
$\kappa > 0$ is used to control the simulated photon count,
i.e., the noise level. We will also denote the vector
$b^\dagger$ as the one with ideal data: $b^\dagger_i =
\kappa \mathcal R[ f ]( \theta_{v_i}, t_{r_i} )$.
In these studies, we have used $v = \thetasize$ and
$r = \tsize$ in order to reconstruct images consisting of
$\isize \times \jsize$ pixels.

We also experiment using data from a human cardiac SPECT
study. In this set of experiments, the data
set had $v = 60$ and $r = 64$ and we reconstructed images
with $64 \times 64$ pixels. No attenuation correction has
been applied to the data.

\paragraph{String Selection and Weights}

We denote by SAEM-$T$ an SAEM algorithm with $T$ strings,
and the strings were built by randomly shuffling the data
and then partitioning it in $T$ contiguous
chunks of uniform size, each of which would become
a string. It was done
like that for two reasons. First, it is often recognized
that a random order performs as well, if not
better, than most other of practical viability
in sequential algorithms. Second,
this avoids unfair comparisons that could arise from
specific choices of subsets, which could possibly be more
suitable to one EM variant than to others.

We wanted to keep the classical maximum likelihood model, so
we used non-weighted averaging, that is $\omega_i = 1 / T$.
More sophisticated weight selection schemes are conceivable,
but this subject is beyond the scope of the present paper.

\paragraph{Scaling Matrix}

In the algorithms we use, the scaling matrix $D( x )$ is a
diagonal matrix with non-zero elements given by $x_j / p_j$
for a fixed, but unspecified, set of weights $p_j > 0$. We have
chosen $p_j = \sum_{i = 1}^n a_{ij}$. Several other
forms where possible, including some dependent on the
particular method being used, but we chose to keep the
experiments understandable and to have a uniform treatment
among the algorithms, rather than to complicate
the results introducing another varying parameter.

\paragraph{Stepsize}

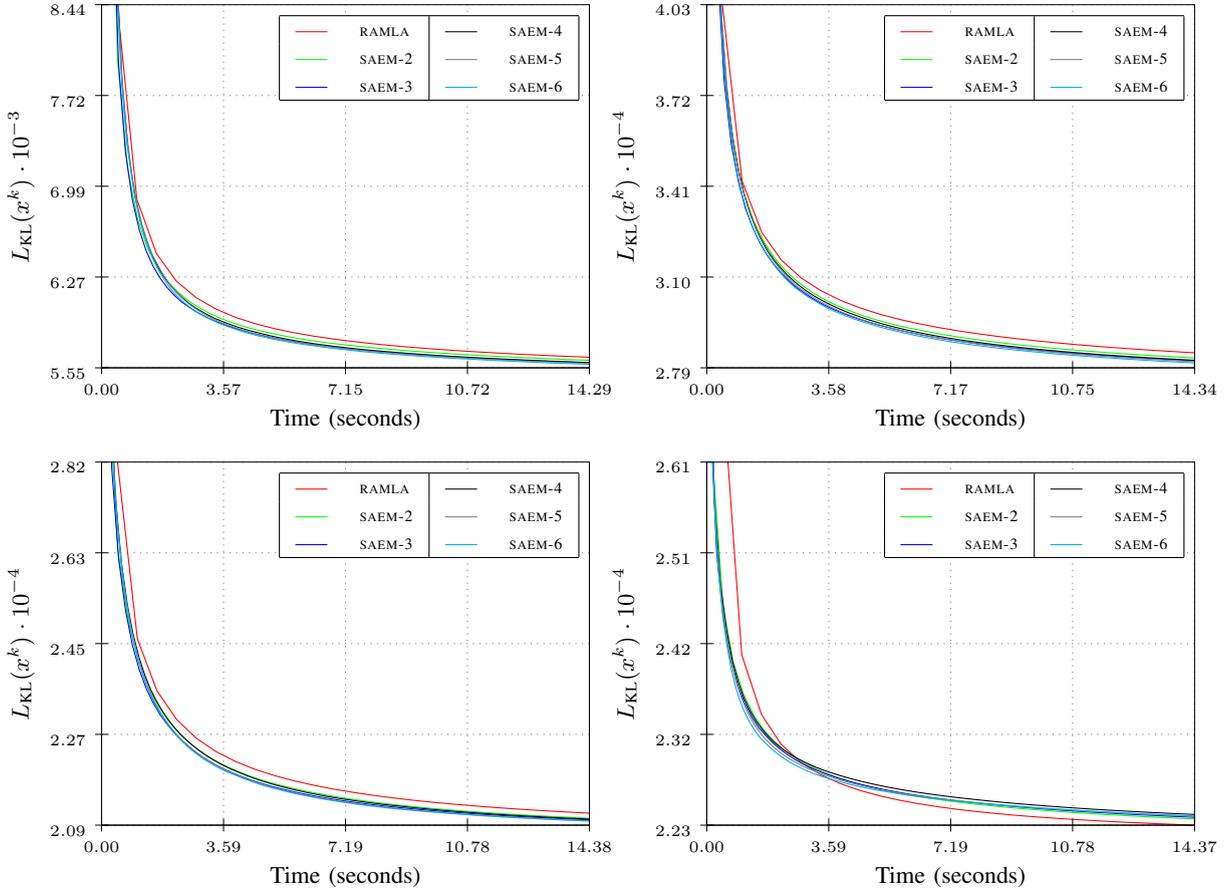
\begin{figure}
   \setlength{\grftotalwidth}{0.5\textwidth}%
   \begin{center}
      \input{expnn_plot_data.tex}%
      \begin{grfgraphic*}[0.75]{%
         \setxaxis%
         \setyaxis%
         \grfwindow%
         \grfylabel[L]{\footnotesize$L_{\text{KL}}( x^k ) \cdot 10^{-\yexp}$}%
         \grfxlabel{\footnotesize Time (seconds)}%
      }
         \setxgrid%
         \setygrid%
         \draw[red] plot file {expnn_saem-1_plot_timexll.data};%
         \draw[green] plot file {expnn_saem-2_plot_timexll.data};%
         \draw[blue] plot file {expnn_saem-3_plot_timexll.data};%
         \draw[black] plot file {expnn_saem-4_plot_timexll.data};%
         \draw[gray] plot file {expnn_saem-5_plot_timexll.data};%
         \draw[cyan] plot file {expnn_saem-6_plot_timexll.data};%
         {\tiny%
            \grftablegend[tr]{|rl|rl|}{%
               \hline
               \grflegsymbol[2em]{\grflegline[red]}&\textsc{ramla}&
               \grflegsymbol[2em]{\grflegline[black]}&\textsc{saem-4}\\
               \grflegsymbol[2em]{\grflegline[green]}&\textsc{saem-2}&
               \grflegsymbol[2em]{\grflegline[gray]}&\textsc{saem-5}\\
               \grflegsymbol[2em]{\grflegline[blue]}&\textsc{saem-3}&
               \grflegsymbol[2em]{\grflegline[cyan]}&\textsc{saem-6}\\
               \hline
            }%
         }%
      \end{grfgraphic*}%
      \input{exp1000_plot_data.tex}%
      \begin{grfgraphic*}[0.75]{%
         \setxaxis%
         \setyaxis%
         \grfwindow%
         \grfylabel[L]{\footnotesize$L_{\text{KL}}( x^k ) \cdot 10^{-\yexp}$}%
         \grfxlabel{\footnotesize Time (seconds)}%
      }
         \setxgrid%
         \setygrid%
         \draw[red] plot file {exp1000_saem-1_plot_timexll.data};%
         \draw[green] plot file {exp1000_saem-2_plot_timexll.data};%
         \draw[blue] plot file {exp1000_saem-3_plot_timexll.data};%
         \draw[black] plot file {exp1000_saem-4_plot_timexll.data};%
         \draw[gray] plot file {exp1000_saem-5_plot_timexll.data};%
         \draw[cyan] plot file {exp1000_saem-6_plot_timexll.data};%
         {\tiny%
            \grftablegend[tr]{|rl|rl|}{%
               \hline
               \grflegsymbol[2em]{\grflegline[red]}&\textsc{ramla}&
               \grflegsymbol[2em]{\grflegline[black]}&\textsc{saem-4}\\
               \grflegsymbol[2em]{\grflegline[green]}&\textsc{saem-2}&
               \grflegsymbol[2em]{\grflegline[gray]}&\textsc{saem-5}\\
               \grflegsymbol[2em]{\grflegline[blue]}&\textsc{saem-3}&
               \grflegsymbol[2em]{\grflegline[cyan]}&\textsc{saem-6}\\
               \hline
            }%
         }%
      \end{grfgraphic*}\vspace{1em}
      \input{exp500_plot_data.tex}%
      \begin{grfgraphic*}[0.75]{%
         \setxaxis%
         \setyaxis%
         \grfwindow%
         \grfylabel[L]{\footnotesize$L_{\text{KL}}( x^k ) \cdot 10^{-\yexp}$}%
         \grfxlabel{\footnotesize Time (seconds)}%
      }
         \setxgrid%
         \setygrid%
         \draw[red] plot file {exp500_saem-1_plot_timexll.data};%
         \draw[green] plot file {exp500_saem-2_plot_timexll.data};%
         \draw[blue] plot file {exp500_saem-3_plot_timexll.data};%
         \draw[black] plot file {exp500_saem-4_plot_timexll.data};%
         \draw[gray] plot file {exp500_saem-5_plot_timexll.data};%
         \draw[cyan] plot file {exp500_saem-6_plot_timexll.data};%
         {\tiny%
            \grftablegend[tr]{|rl|rl|}{%
               \hline
               \grflegsymbol[2em]{\grflegline[red]}&\textsc{ramla}&
               \grflegsymbol[2em]{\grflegline[black]}&\textsc{saem-4}\\
               \grflegsymbol[2em]{\grflegline[green]}&\textsc{saem-2}&
               \grflegsymbol[2em]{\grflegline[gray]}&\textsc{saem-5}\\
               \grflegsymbol[2em]{\grflegline[blue]}&\textsc{saem-3}&
               \grflegsymbol[2em]{\grflegline[cyan]}&\textsc{saem-6}\\
               \hline
            }%
         }%
      \end{grfgraphic*}%
      \input{exp50_plot_data.tex}%
      \begin{grfgraphic*}[0.75]{%
         \setxaxis%
         \setyaxis%
         \grfwindow%
         \grfylabel[L]{\footnotesize$L_{\text{KL}}( x^k ) \cdot 10^{-\yexp}$}%
         \grfxlabel{\footnotesize Time (seconds)}%
      }
         \setxgrid%
         \setygrid%
         \draw[red] plot file {exp50_saem-1_plot_timexll.data};%
         \draw[green] plot file {exp50_saem-2_plot_timexll.data};%
         \draw[blue] plot file {exp50_saem-3_plot_timexll.data};%
         \draw[black] plot file {exp50_saem-4_plot_timexll.data};%
         \draw[gray] plot file {exp50_saem-5_plot_timexll.data};%
         \draw[cyan] plot file {exp50_saem-6_plot_timexll.data};%
         {\tiny%
            \grftablegend[tr]{|rl|rl|}{%
               \hline
               \grflegsymbol[2em]{\grflegline[red]}&\textsc{ramla}&
               \grflegsymbol[2em]{\grflegline[black]}&\textsc{saem-4}\\
               \grflegsymbol[2em]{\grflegline[green]}&\textsc{saem-2}&
               \grflegsymbol[2em]{\grflegline[gray]}&\textsc{saem-5}\\
               \grflegsymbol[2em]{\grflegline[blue]}&\textsc{saem-3}&
               \grflegsymbol[2em]{\grflegline[cyan]}&\textsc{saem-6}\\
               \hline
            }%
         }%
      \end{grfgraphic*}%
   \end{center}%
   \input{expnn_noiselevel.tex}%
   \edef\nlevela{\noiselevel}%
   \input{exp1000_noiselevel.tex}%
   \edef\nlevelb{\noiselevel}%
   \input{exp500_noiselevel.tex}%
   \edef\nlevelc{\noiselevel}%
   \input{exp50_noiselevel.tex}%
   \edef\nleveld{\noiselevel}%
   \caption{Convergence speed of several SAEM variations
   under different noise condition. From top left in
   clockwise direction:  $\nlevela$\%, $\nlevelb$\%,
   $\nlevelc$\% and $\nleveld$\% of relative noise $\| b -
   b^\dagger \| / \| b^\dagger \|$.
   Notice that if up to $T_{\max}$ strings can be run in parallel,
   then at any given time point, SAEM-$T$ has similar log-likelihood
   values for $T = 1, \dots, T_{\max}$. Running time was
   estimated by accumulating high-definition clock information
   of iteration computation, ignoring input/output time.}
   \label{fig:timexll}
\end{figure}

We are required to determine the nonsummable stepsize
sequence $\{ \lambda_k \} \subset \mathbf R_{+}$. We use
the rule, for SAEM-$T$:
\begin{equation}
   \lambda_k = \frac{\lambda_0^T}{k^{0.51} / T + 1},
\end{equation}
where $\lambda_0^T$ was selected as the
largest value for which
SAEM-$T$ would not lead to negative values in the first
iteration. This technique
for the starting parameter leads to the fastest possible
convergence in terms of objective function decrease, and
the scaling by the inverse of the number of strings accounts
for the $1/T$ factor, caused by the averaging,
in~\eqref{eq:iterlambdaoverT}. Figure~\ref{fig:timexll}
shows that this is indeed a well balanced
parameter rule, because it leads all SAEM variations to have
similar log-likelihood values at any time point, as long as
all the strings can be run simultaneously in hardware. It
can also be seen that this behavior remains consistent under
several noise regimes. This same kind of plot is shown in the
left-hand side of Figure~\ref{fig:llxtvspect} for the, smaller,
real data example, where the comparison is no longer as
balanced, because parallelism overhead does not pay off with
small images, but in this case computation time is negligible
anyway.

\paragraph{Starting Image}

Initial guess for the algorithms was a uniform image with an
expected photon count equal to the data. That is, $x^0_j =
\alpha$ where $\alpha$ is such that
\begin{equation}
   \sum_{i = 0}^m( A x^0 )_i = \sum_{i = 0}^mb_i.
\end{equation}
The value of $\alpha$ can be easily obtained as $\alpha =
\sum_{i = 0}^mb_i / \sum_{i = 0}^m( A \mathbf 1 )_i$,
where $\mathbf 1$ is the vector whose components are all
equal to $1$.

\subsection{Image Reconstruction Analysis}\label{subsec:ira}

Figure~\ref{fig:timexll} shows that all SAEM-$T$ algorithms
present similar convergence speeds when measured as the ratio of
objective function decrease and computation time. Notice that
at iteration $k$, SAEM-$T$ has a smaller log-likelihood value than
SAEM-$( T + 1 )$. Hence, more strings means slower algorithms
iteration-wise because there is less incrementalism present.
It is widely recognized, however, that log-likelihood cannot be
used as an image quality measure because over-fitting the image
to the data will amplify high frequency components of the noise.
This leads to the need for other ways of assessing reconstruction
quality.

In the simulated studies, we make use of the mean squared error,
which requires full knowledge about the ideal sought-after image:
\begin{equation}
   MSE( x ) := \frac{\| x - x^\dagger \|_2^2}{ \| x^\dagger \|_2^2 },
\end{equation}
where $x^\dagger$ is the noise-free discretization of the
Shepp-Logan head phantom, properly scaled by $\kappa$. Another
functional we have used to measure image quality was the
Total Variation:
\begin{equation*}
   TV( x ) := \sum_{i = 1}^{r_1}\sum_{j = 1}^{r_2}\sqrt{
   ( x_{i, j} - x_{i, j - 1} )^2 + ( x_{i, j} - x_{i - 1, j} )^2 },
\end{equation*}
where $r_1r_2 = n$ and we use the convention $x_{i, j} =
x_{i( r_2 - 1 ) + j}$ for $i \in \{ 1, 2, \dots, r_1 \}$ and
$j \in \{ 1, 2, \dots, r_2 \}$ with the boundary condition
$x_{0, k} = x_{ k, 0 } = 0$. Unlike the $MSE$, the Total
Variation alone cannot be used as a measure of image quality.
However, given two images of piecewise constant emission
rates with the same Poisson likelihood, the one with the
smaller Total Variation is more likely to have better image
quality. Since neither the log-likelihood nor the Total
Variation require knowledge of the ideal image, the combination
of both can be used also in the evaluation of the SPECT
reconstruction.

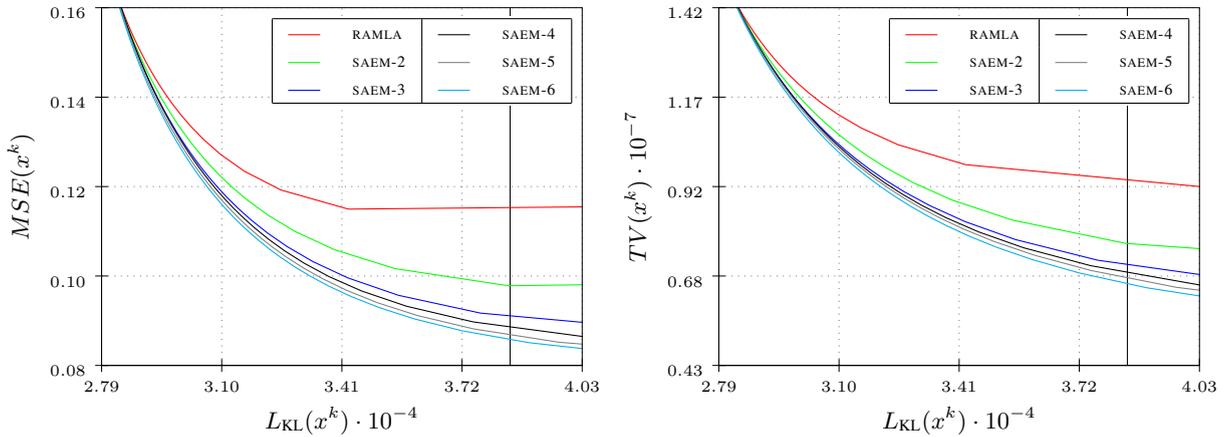
\begin{figure}
   \input{exp1000_llpoint.tex}%
   \input{exp1000_plot_data.tex}%
   \setlength{\grftotalwidth}{0.49\textwidth}%
   \begin{center}
      \begin{grfgraphic*}[0.75]{%
         \setlxaxis%
         \setmseyaxis%
         \grfwindow%
         \grfxlabel{\footnotesize$L_{\text{KL}}( x^k ) \cdot 10^{-\yexp}$}%
         \grfylabel[L]{\footnotesize $MSE( x^k )$}%
      }
         \draw[grfaxisstyle] (\llpoint,0) -- (\llpoint,1);%
         \setxgrid%
         \setygrid%
         {\tiny%
            \grftablegend[tr]{|rl|rl|}{%
               \hline
               \grflegsymbol[2em]{\grflegline[red]}&\textsc{ramla}&
               \grflegsymbol[2em]{\grflegline[black]}&\textsc{saem-4}\\
               \grflegsymbol[2em]{\grflegline[green]}&\textsc{saem-2}&
               \grflegsymbol[2em]{\grflegline[gray]}&\textsc{saem-5}\\
               \grflegsymbol[2em]{\grflegline[blue]}&\textsc{saem-3}&
               \grflegsymbol[2em]{\grflegline[cyan]}&\textsc{saem-6}\\
               \hline
            }%
         }%
         \draw[red] plot file {exp1000_saem-1_plot_llxmse.data};%
         \draw[green] plot file {exp1000_saem-2_plot_llxmse.data};%
         \draw[blue] plot file {exp1000_saem-3_plot_llxmse.data};%
         \draw[black] plot file {exp1000_saem-4_plot_llxmse.data};%
         \draw[gray] plot file {exp1000_saem-5_plot_llxmse.data};%
         \draw[cyan] plot file {exp1000_saem-6_plot_llxmse.data};%
      \end{grfgraphic*}\hfill%
      \begin{grfgraphic*}[0.75]{%
         \setlxaxis%
         \settvyaxis%
         \grfwindow%
         \grfxlabel{\footnotesize$L_{\text{KL}}( x^k ) \cdot 10^{-\yexp}$}%
         \grfylabel[L]{\footnotesize $TV( x^k ) \cdot 10^{-\tvexp}$}%
      }
         \setxgrid%
         \setygrid%
         \draw[grfaxisstyle] (\llpoint,0) -- (\llpoint,1);%
         {\tiny%
            \grftablegend[tr]{|rl|rl|}{%
               \hline
               \grflegsymbol[2em]{\grflegline[red]}&\textsc{ramla}&
               \grflegsymbol[2em]{\grflegline[black]}&\textsc{saem-4}\\
               \grflegsymbol[2em]{\grflegline[green]}&\textsc{saem-2}&
               \grflegsymbol[2em]{\grflegline[gray]}&\textsc{saem-5}\\
               \grflegsymbol[2em]{\grflegline[blue]}&\textsc{saem-3}&
               \grflegsymbol[2em]{\grflegline[cyan]}&\textsc{saem-6}\\
               \hline
            }%
         }%
         \draw[red] plot file {exp1000_saem-1_plot_llxtv.data};%
         \draw[green] plot file {exp1000_saem-2_plot_llxtv.data};%
         \draw[blue] plot file {exp1000_saem-3_plot_llxtv.data};%
         \draw[black] plot file {exp1000_saem-4_plot_llxtv.data};%
         \draw[gray] plot file {exp1000_saem-5_plot_llxtv.data};%
         \draw[cyan] plot file {exp1000_saem-6_plot_llxtv.data};%
      \end{grfgraphic*}
   \end{center}
   \caption{Relative squared error and Total Variation as functions
   of the log-likelihood for iterations of SAEM variants.
   Notice that both the relative error and the Total Variation
   values, for a fixed likelihood level, are
   decreasing functions of the number of strings. The solid
   vertical lines show the log-likelihood level of the images
   of Figure~\ref{fig:images}.}%
   \label{fig:llxmse}%
\end{figure}

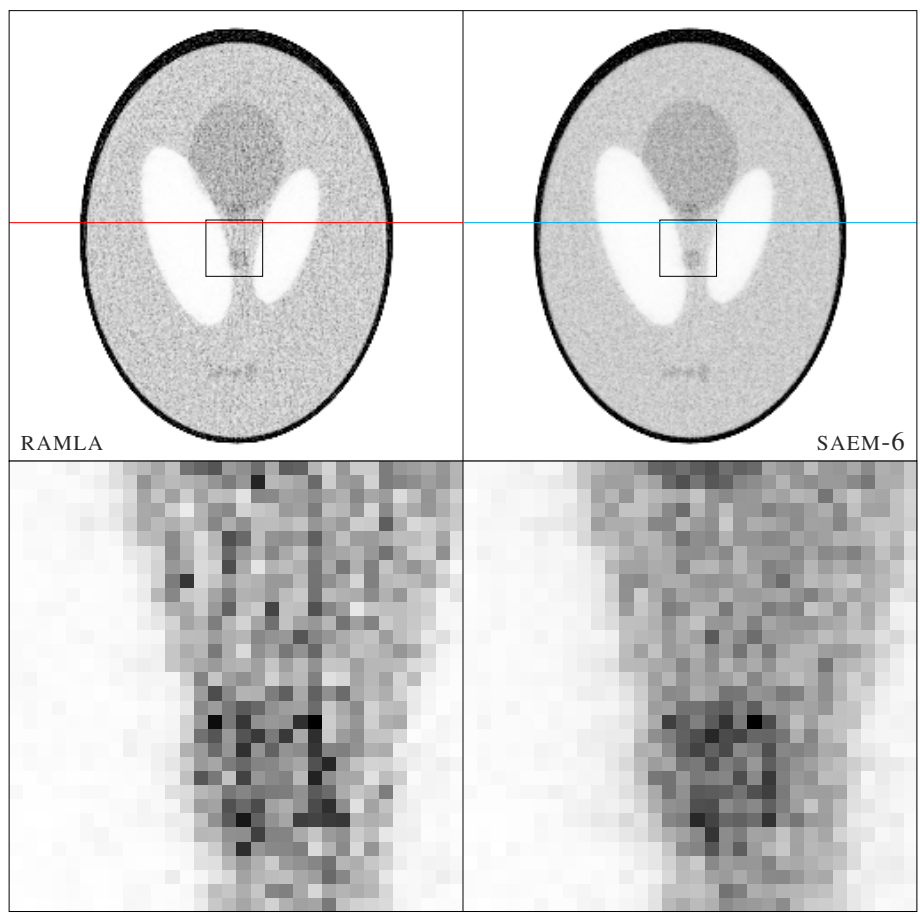
\begin{figure}
   \input{exp1000_sub_data.tex}%
   \setlength{\grftotalwidth}{0.75\textwidth}%
   \begin{center}%
      \begin{grfgraphic}{%
         \def\grfymin{-1}%
         \def\grfymax{1}%
         \def\grfxmin{-2}%
         \def\grfxmax{2}%
         \grfwindow%
      }
         \node[inner sep=0pt] at (-1,0) {\includegraphics[width=2\grfxunit]{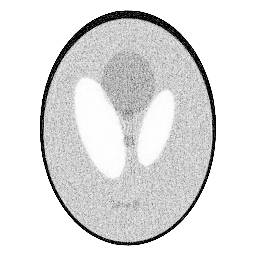}};%
         \node[inner sep=0pt] at (1,0) {\includegraphics[width=2\grfxunit]{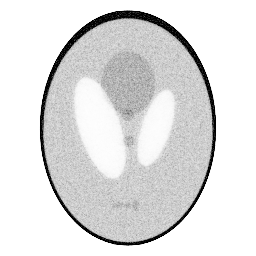}};%
         \draw[grfaxisstyle] (0,-1) -- (0,1);%
         \node[fill=white,anchor=south west,opacity=0.85] at (-2,-1) {\small\textsc{ramla\vphantom{6}}};%
         \node[fill=white,anchor=south east,opacity=0.85] at (2,-1) {\small\textsc{saem-6}};%
         \begin{scope}[xshift=-1\grfxunit]%
            \subrect%
            \draw[red] (-1,\profy) -- (1,\profy);%
         \end{scope}%
         \begin{scope}[xshift=1\grfxunit]%
            \subrect%
            \draw[cyan] (-1,\profy) -- (1,\profy);%
         \end{scope}%
      \end{grfgraphic}
      \begin{grfgraphic}{%
         \def\grfymin{-1}%
         \def\grfymax{1}%
         \def\grfxmin{-2}%
         \def\grfxmax{2}%
         \grfwindow%
      }
         \node[inner sep=0pt] at (-1,0) {\includegraphics[width=2\grfxunit]{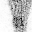}};%
         \node[inner sep=0pt] at (1,0) {\includegraphics[width=2\grfxunit]{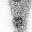}};%
         \draw[grfaxisstyle] (0,-1) -- (0,1);%
      \end{grfgraphic}%
   \end{center}%
   \caption{Details of reconstruction obtained by RAMLA (left)
   and SAEM-6 (right). Both images where these details taken
   from have, up to linear interpolation errors, the
   log-likelihood level indicated by the solid vertical
   lines in Figure~\ref{fig:llxmse}. It is noticeable
   that the smaller Total Variance is
   reflected in both smoother large areas and better
   resolved sharp transitions. The horizontal solid colored
   lines show where the profiles of Figure~\ref{fig:profiles}
   where taken from.}%
   \label{fig:images}%
\end{figure}

\begin{figure}
   \input{exp1000_plot_data.tex}%
   \setlength{\grftotalwidth}{\textwidth}%
   \begin{center}%
      \begin{grfgraphic*}[0.5]{%
         \grfyaxis{[]0;0.25;0.5;0.75;1[]}{[]\tiny$0$;\tiny$0.25$;\tiny$0.5$;\tiny$0.75$;\tiny$1$[]}%
         \grfxaxis{[]-1;-0.5;0;0.5;1[]}{[]\tiny$-1$;\tiny$-0.5$;\tiny$0$;\tiny$0.5$;\tiny$1$[]}%
         \def\grfxmax{1.05}%
      }
         \def\grfxmax{1}%
         \grfygrid[dotted,gray]{[]0;0.25;0.5;0.75;1[]}{[]\tiny$0$;\tiny$0.25$;\tiny$0.5$;\tiny$0.75$;\tiny$1$[]}%
         \grfxgrid[dotted,gray]{[]-1;-0.5;0;0.5;1[]}{[]\tiny$-1$;\tiny$-0.5$;\tiny$0$;\tiny$0.5$;\tiny$1$[]}%
         \draw[red] plot file {exp1000_saem-1_profile.data};%
         \draw[cyan] plot file {exp1000_saem-6_profile.data};%
         \draw plot file {profile_dagger.data};%
         \def\grfxmax{1.05}%
         {\tiny%
            \grftablegend[tr]{|rl|}{%
               \hline
               \grflegsymbol[2em]{\grflegline[red]}&\textsc{ramla}\\
               \grflegsymbol[2em]{\grflegline[cyan]}&\textsc{saem-6}\\
               \grflegsymbol[2em]{\grflegline}&\textsc{True image}\\
               \hline
            }%
         }%
      \end{grfgraphic*}%
   \end{center}%
   \caption{Profile lines from images in Figure~\ref{fig:images}. Notice how SAEM-6 reconstruction presents
   less overshoot and more smoothness than RAMLA reconstruction.}%
   \label{fig:profiles}%
\end{figure}
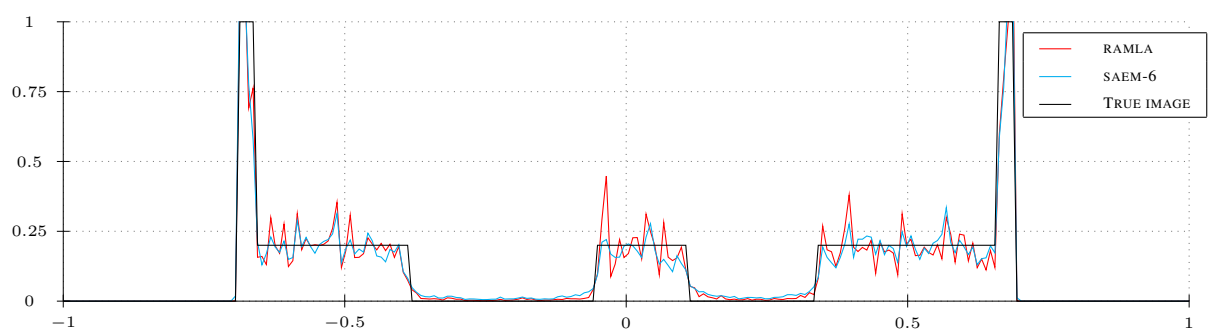

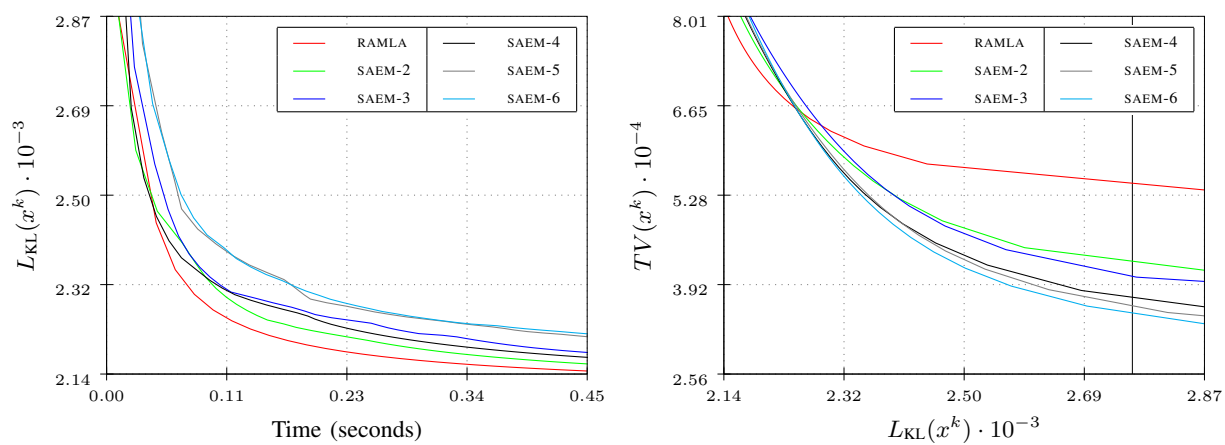
\begin{figure}
   \input{spect_llpoint.tex}%
   \input{spect_plot_data.tex}%
   \setlength{\grftotalwidth}{0.49\textwidth}%
   \begin{center}%
      \begin{grfgraphic*}[0.75]{%
         \setxaxis%
         \setyaxis%
         \grfwindow%
         \grfylabel[L]{\footnotesize$L_{\text{KL}}( x^k ) \cdot 10^{-\yexp}$}%
         \grfxlabel{\footnotesize Time (seconds)\vphantom{$L_{\text{KL}}( x^k ) \cdot 10^{-\yexp}$}}%
      }
         \setxgrid%
         \setygrid%
         \draw[red] plot file {spect_saem-1_plot_timexll.data};%
         \draw[green] plot file {spect_saem-2_plot_timexll.data};%
         \draw[blue] plot file {spect_saem-3_plot_timexll.data};%
         \draw[black] plot file {spect_saem-4_plot_timexll.data};%
         \draw[gray] plot file {spect_saem-5_plot_timexll.data};%
         \draw[cyan] plot file {spect_saem-6_plot_timexll.data};%
         {\tiny%
            \grftablegend[tr]{|rl|rl|}{%
               \hline
               \grflegsymbol[2em]{\grflegline[red]}&\textsc{ramla}&
               \grflegsymbol[2em]{\grflegline[black]}&\textsc{saem-4}\\
               \grflegsymbol[2em]{\grflegline[green]}&\textsc{saem-2}&
               \grflegsymbol[2em]{\grflegline[gray]}&\textsc{saem-5}\\
               \grflegsymbol[2em]{\grflegline[blue]}&\textsc{saem-3}&
               \grflegsymbol[2em]{\grflegline[cyan]}&\textsc{saem-6}\\
               \hline
            }%
         }%
      \end{grfgraphic*}\hfill%
      \begin{grfgraphic*}[0.75]{%
         \setlxaxis%
         \settvyaxis%
         \grfwindow%
         \grfxlabel{\footnotesize$L_{\text{KL}}( x^k ) \cdot 10^{-\yexp}$}%
         \grfylabel[L]{\footnotesize $TV( x^k ) \cdot 10^{-\tvexp}$}%
      }
         \setxgrid%
         \setygrid%
         \draw[grfaxisstyle] (\llpoint,0) -- (\llpoint,1);%
         {\tiny%
            \grftablegend[tr]{|rl|rl|}{%
               \hline
               \grflegsymbol[2em]{\grflegline[red]}&\textsc{ramla}&
               \grflegsymbol[2em]{\grflegline[black]}&\textsc{saem-4}\\
               \grflegsymbol[2em]{\grflegline[green]}&\textsc{saem-2}&
               \grflegsymbol[2em]{\grflegline[gray]}&\textsc{saem-5}\\
               \grflegsymbol[2em]{\grflegline[blue]}&\textsc{saem-3}&
               \grflegsymbol[2em]{\grflegline[cyan]}&\textsc{saem-6}\\
               \hline
            }%
         }%
         \draw[red] plot file {spect_saem-1_plot_llxtv.data};%
         \draw[green] plot file {spect_saem-2_plot_llxtv.data};%
         \draw[blue] plot file {spect_saem-3_plot_llxtv.data};%
         \draw[black] plot file {spect_saem-4_plot_llxtv.data};%
         \draw[gray] plot file {spect_saem-5_plot_llxtv.data};%
         \draw[cyan] plot file {spect_saem-6_plot_llxtv.data};%
      \end{grfgraphic*}%
   \end{center}%
   \caption{Speed convergence and Total Variation results
   for the human SPECT study. The solid line in
   the graphic at the right indicates the log-likelihood
   level of the images in Figure~\ref{fig:imagesspect}.}%
   \label{fig:llxtvspect}%
\end{figure}

\begin{figure}
   \input{spect_sub_data.tex}%
   \setlength{\grftotalwidth}{0.75\textwidth}%
   \begin{center}%
      \begin{grfgraphic}{%
         \def\grfymin{-1}%
         \def\grfymax{1}%
         \def\grfxmin{-2}%
         \def\grfxmax{2}%
         \grfwindow%
      }
         \node[inner sep=0pt] at (-1,0) {\includegraphics[width=2\grfxunit]{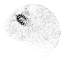}};%
         \node[inner sep=0pt] at (1,0) {\includegraphics[width=2\grfxunit]{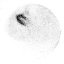}};%
         \draw[grfaxisstyle] (0,-1) -- (0,1);%
         \node[fill=white,anchor=south west,opacity=0.85] at (-2,-1) {\small\textsc{ramla\vphantom{6}}};%
         \node[fill=white,anchor=south east,opacity=0.85] at (2,-1) {\small\textsc{saem-6}};%
         \begin{scope}[xshift=-1\grfxunit]%
            \subrect%
         \end{scope}%
         \begin{scope}[xshift=1\grfxunit]%
            \subrect%
         \end{scope}%
      \end{grfgraphic}
      \begin{grfgraphic}{%
         \def\grfymin{-1}%
         \def\grfymax{1}%
         \def\grfxmin{-2}%
         \def\grfxmax{2}%
         \grfwindow%
      }
         \node[inner sep=0pt] at (-1,0) {\includegraphics[width=2\grfxunit]{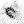}};%
         \node[inner sep=0pt] at (1,0) {\includegraphics[width=2\grfxunit]{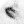}};%
         \draw[grfaxisstyle] (0,-1) -- (0,1);%
      \end{grfgraphic}%
   \end{center}%
   \caption{Male human cardiac SPECT study. The log-likelihood
   level of the images is indicated by the solid lines in
   the graphic at the right in Figure~\ref{fig:llxtvspect}.
   Data kindly provided by Roberto Isoardi -- Fundaci\'on
   Escuela de Medicina Nuclear (FUESMEN) -- Mendoza,
   Argentina}%
   \label{fig:imagesspect}%
\end{figure}

Figure~\ref{fig:llxmse} shows plots of both figures of merit
as functions of the log-likelihood for one of the simulated
experiments. On the left we have $MSE$ and on the right $TV$.
The remarkable feature of these plots is the fact that the
quality of the image obtained by SAEM-$T$, at any given fixed
likelihood level is an increasing function of $T$, the number
of strings. This behavior is unaltered as we vary the
noise level. Notice the same effect in the $TV$ versus
$L_{\text{KL}}$ curve shown in the graphic at the right
of Figure~\ref{fig:llxtvspect}, and that the difference in the
Total Variation translates into significant visual improvement
in the real data study (Figure~\ref{fig:imagesspect}), even
more than it does in the simulated case (Figures~\ref{fig:images}~and~\ref{fig:profiles}).

\section{Conclusions}\label{sec:conclusions}

We have presented the new String-Averaging Expectation-Maximization
family of algorithms. The theoretical convergence properties
of the method were studied and experimental
evidence was provided of the technique's suitability for
good quality reconstruction in realistic settings.

The theory we have developed here can handle more general
regularized objective functions, as they
can achieve a better balance between smoothness and adherence
to the data than traditional techniques currently used in
tomographic scanners.

Future work could focus on comparative evaluation of SAEM
against other paradigms, such as OS-EM or block-RAMLA, in light of
the $TV$ versus $L_{\text{KL}}$ plots, which we have shown to be
useful to assess image quality. It is potentially worth
investigating \textit{a posteriori} stopping criteria based
on such plots, in the spirit of the L-curve~\cite{hansen}.


\textbf{Acknowledgments}
We are sincerely grateful to the anonymous reviewers
whose insightful comments encouraged us to extend some of the
experimental work in the paper. This resulted in discovering further
advantages of the proposed SAEM algorithm in the
final version.

Work by E. H. was partially supported by FAPESP grant
2013/16508-3, Brazil.
Y.~C. was partially
supported by the United States-Israel Binational Science Foundation (BSF)
grant number 200912 and by US Department of Army award number
W81XWH-10-1-0170.
T.~B.  was partially supported by the National Science Council of the Republic of China,
Taiwan (NSC 97-2118-M-214-001-MY2).
I.~C. was partially supported by the National Center for Theoretical
Sciences (Taipei Office) and the National Science Council
of the Republic of China (99-2115-M-002-003-MY3).
A.~D was partially supported
by CNPq grant No 301064/2009-1, Brazil.
M.~J. was partially supported by the National Basic Research and
Development Program of China (973 Program) (2011CB809105), National
Science Foundation of China (61121002, 10990013, 60325101).
H.~L. was supported by grants from National Science Council, National Center
for Theoretical Sciences, Center of Mathematical Modeling and
Scientific Computing at National Chiao Tung University in Taiwan.
\bigskip

\def\cprime{$'$}

\end{document}

%% file: sl.tex
\newcommand{\shepplogan}
{%
   \begin{scope}[line width=0pt]
      \path[color=black!100,draw,fill] (0,0)       ellipse (0.69 and 0.92);
      \path[color=black!20,draw,fill]  (0,-0.0184) ellipse (0.6624 and 0.874);

      \path[color=black!30,draw,fill] (0,0.35)       ellipse (0.21 and 0.25);
      \path[color=black!30,draw,fill] (0,-0.1)       ellipse (0.046 and 0.046);
      \path[color=black!30,draw,fill] (-0.08,-0.605) ellipse (0.046 and 0.023);
      \path[color=black!30,draw,fill] (0,-0.606)     ellipse (0.023 and 0.023);
      \path[color=black!30,draw,fill] (0.06,-0.605)  ellipse (0.023 and 0.046);
      \path[color=black!30,draw,fill] (0.06,-0.605)  ellipse (0.023 and 0.046);

      \path[color=black!0,draw,fill,xshift=0.22\grfxunit,rotate=-18] (0,0) ellipse (0.11 and 0.31);
      \path[color=black!0,draw,fill,xshift=-0.22\grfxunit,rotate=18] (0,0) ellipse (0.16 and 0.41);
      \begin{scope}
         \path[clip,xshift=-0.22\grfxunit,rotate=18] (0,0)    ellipse (0.16 and 0.41);
         \path[color=black!10,draw,fill]             (0,0.35) ellipse (0.21 and 0.25);
         \path[color=black!10,draw,fill]             (0,-0.1) ellipse (0.046 and 0.046);
      \end{scope}
   \end{scope}

   \path[color=black!30,draw,fill]    (0,0.1)  ellipse (0.046 and 0.046);
   \begin{scope}
      \path[clip]                     (0,0.1)  ellipse (0.046 and 0.046);
      \path[color=black!40,draw,fill] (0,0.35) ellipse (0.21 and 0.25);
   \end{scope}
}

%% file: exp1000_sub_data.tex
\def\subrect{\draw[grfaxisstyle](-0.1328,0.0703) rectangle (0.1172,-0.1797);}%
\def\isize{256}%
\def\jsize{256}%
\def\thetasize{288}%
\def\tsize{256}%
\def\profy{0.0585937500000000}%
\def\subprofy{\relax}%
\def\profymax{2000}%

%% file: figradonteta.tex
\def\tetarad{1.04719755119659774560}
\def\teta{60}
\def\tetameio{30}
 

%% file: expnn_plot_data.tex
\def\maxtmp{14.291}%
\def\mintmp{0.119}%
\def\minlllabel{5.5488174 \cdot 10^{3}}%
\def\maxlllabel{8.4396115 \cdot 10^{3}}%
\def\setxaxis{\grfxaxis[dotted]{[]0.00;0.25;0.50;0.75;1.00[]}{[]\tiny$0.00$;\tiny$3.57$;\tiny$7.15$;\tiny$10.72$;\tiny$14.29$[]}}%
\def\setxgrid{\grfxgrid[dotted,gray]{[]0.00;0.25;0.50;0.75;1.00[]}{[]0.00;0.25;0.50;0.75;1.00[]}}%
\def\setyaxis{\grfyaxis[dotted]{[]0.00;0.25;0.50;0.75;1.00[]}{[]\tiny$5.55$;\tiny$6.27$;\tiny$6.99$;\tiny$7.72$;\tiny$8.44$[]}}%
\def\setlxaxis{\grfxaxis[dotted]{[]0.00;0.25;0.50;0.75;1.00[]}{[]\tiny$5.55$;\tiny$6.27$;\tiny$6.99$;\tiny$7.72$;\tiny$8.44$[]}}%
\def\setygrid{\grfygrid[dotted,gray]{[]0.00;0.25;0.50;0.75;1.00[]}{[]0.00;0.25;0.50;0.75;1.00[]}}%
\def\yexp{3}%
\def\settvyaxis{\grfyaxis[dotted]{[]0.00;0.25;0.50;0.75;1.00[]}{[]\tiny$2.01$;\tiny$2.43$;\tiny$2.86$;\tiny$3.29$;\tiny$3.72$[]}}%
\def\tvexp{6}%
\def\setmseyaxis{\grfyaxis[dotted]{[]0.00;0.25;0.50;0.75;1.00[]}{[]\tiny$0.07$;\tiny$0.07$;\tiny$0.08$;\tiny$0.09$;\tiny$0.09$[]}}%

%% file: exp1000_plot_data.tex
\def\maxtmp{14.338}%
\def\mintmp{0.120}%
\def\minlllabel{2.7886208 \cdot 10^{4}}%
\def\maxlllabel{4.0282057 \cdot 10^{4}}%
\def\setxaxis{\grfxaxis[dotted]{[]0.00;0.25;0.50;0.75;1.00[]}{[]\tiny$0.00$;\tiny$3.58$;\tiny$7.17$;\tiny$10.75$;\tiny$14.34$[]}}%
\def\setxgrid{\grfxgrid[dotted,gray]{[]0.00;0.25;0.50;0.75;1.00[]}{[]0.00;0.25;0.50;0.75;1.00[]}}%
\def\setyaxis{\grfyaxis[dotted]{[]0.00;0.25;0.50;0.75;1.00[]}{[]\tiny$2.79$;\tiny$3.10$;\tiny$3.41$;\tiny$3.72$;\tiny$4.03$[]}}%
\def\setlxaxis{\grfxaxis[dotted]{[]0.00;0.25;0.50;0.75;1.00[]}{[]\tiny$2.79$;\tiny$3.10$;\tiny$3.41$;\tiny$3.72$;\tiny$4.03$[]}}%
\def\setygrid{\grfygrid[dotted,gray]{[]0.00;0.25;0.50;0.75;1.00[]}{[]0.00;0.25;0.50;0.75;1.00[]}}%
\def\yexp{4}%
\def\settvyaxis{\grfyaxis[dotted]{[]0.00;0.25;0.50;0.75;1.00[]}{[]\tiny$0.43$;\tiny$0.68$;\tiny$0.92$;\tiny$1.17$;\tiny$1.42$[]}}%
\def\tvexp{7}%
\def\setmseyaxis{\grfyaxis[dotted]{[]0.00;0.25;0.50;0.75;1.00[]}{[]\tiny$0.08$;\tiny$0.10$;\tiny$0.12$;\tiny$0.14$;\tiny$0.16$[]}}%

%% file: exp500_plot_data.tex
\def\maxtmp{14.380}%
\def\mintmp{0.118}%
\def\minlllabel{2.0877339 \cdot 10^{4}}%
\def\maxlllabel{2.8170202 \cdot 10^{4}}%
\def\setxaxis{\grfxaxis[dotted]{[]0.00;0.25;0.50;0.75;1.00[]}{[]\tiny$0.00$;\tiny$3.59$;\tiny$7.19$;\tiny$10.78$;\tiny$14.38$[]}}%
\def\setxgrid{\grfxgrid[dotted,gray]{[]0.00;0.25;0.50;0.75;1.00[]}{[]0.00;0.25;0.50;0.75;1.00[]}}%
\def\setyaxis{\grfyaxis[dotted]{[]0.00;0.25;0.50;0.75;1.00[]}{[]\tiny$2.09$;\tiny$2.27$;\tiny$2.45$;\tiny$2.63$;\tiny$2.82$[]}}%
\def\setlxaxis{\grfxaxis[dotted]{[]0.00;0.25;0.50;0.75;1.00[]}{[]\tiny$2.09$;\tiny$2.27$;\tiny$2.45$;\tiny$2.63$;\tiny$2.82$[]}}%
\def\setygrid{\grfygrid[dotted,gray]{[]0.00;0.25;0.50;0.75;1.00[]}{[]0.00;0.25;0.50;0.75;1.00[]}}%
\def\yexp{4}%
\def\settvyaxis{\grfyaxis[dotted]{[]0.00;0.25;0.50;0.75;1.00[]}{[]\tiny$1.24$;\tiny$2.26$;\tiny$3.29$;\tiny$4.32$;\tiny$5.34$[]}}%
\def\tvexp{6}%
\def\setmseyaxis{\grfyaxis[dotted]{[]0.00;0.25;0.50;0.75;1.00[]}{[]\tiny$0.09$;\tiny$0.14$;\tiny$0.20$;\tiny$0.25$;\tiny$0.31$[]}}%

%% file: exp50_plot_data.tex
\def\maxtmp{14.373}%
\def\mintmp{0.122}%
\def\minlllabel{2.2262367 \cdot 10^{4}}%
\def\maxlllabel{2.6071394 \cdot 10^{4}}%
\def\setxaxis{\grfxaxis[dotted]{[]0.00;0.25;0.50;0.75;1.00[]}{[]\tiny$0.00$;\tiny$3.59$;\tiny$7.19$;\tiny$10.78$;\tiny$14.37$[]}}%
\def\setxgrid{\grfxgrid[dotted,gray]{[]0.00;0.25;0.50;0.75;1.00[]}{[]0.00;0.25;0.50;0.75;1.00[]}}%
\def\setyaxis{\grfyaxis[dotted]{[]0.00;0.25;0.50;0.75;1.00[]}{[]\tiny$2.23$;\tiny$2.32$;\tiny$2.42$;\tiny$2.51$;\tiny$2.61$[]}}%
\def\setlxaxis{\grfxaxis[dotted]{[]0.00;0.25;0.50;0.75;1.00[]}{[]\tiny$2.23$;\tiny$2.32$;\tiny$2.42$;\tiny$2.51$;\tiny$2.61$[]}}%
\def\setygrid{\grfygrid[dotted,gray]{[]0.00;0.25;0.50;0.75;1.00[]}{[]0.00;0.25;0.50;0.75;1.00[]}}%
\def\yexp{4}%
\def\settvyaxis{\grfyaxis[dotted]{[]0.00;0.25;0.50;0.75;1.00[]}{[]\tiny$2.43$;\tiny$4.04$;\tiny$5.66$;\tiny$7.27$;\tiny$8.89$[]}}%
\def\tvexp{5}%
\def\setmseyaxis{\grfyaxis[dotted]{[]0.00;0.25;0.50;0.75;1.00[]}{[]\tiny$0.17$;\tiny$0.41$;\tiny$0.64$;\tiny$0.87$;\tiny$1.10$[]}}%

%% file: expnn_noiselevel.tex
\def\noiselevel{0.00}%

%% file: exp1000_noiselevel.tex
\def\noiselevel{3.96}%

%% file: exp500_noiselevel.tex
\def\noiselevel{7.94}%

%% file: exp50_noiselevel.tex
\def\noiselevel{25.03}%

%% file: exp1000_llpoint.tex
\def\llpoint{0.85}

%% file: spect_llpoint.tex
\def\llpoint{0.85}

%% file: spect_plot_data.tex
\def\maxtmp{0.450}%
\def\mintmp{0.016}%
\def\minlllabel{2.1392409 \cdot 10^{3}}%
\def\maxlllabel{2.8686833 \cdot 10^{3}}%
\def\setxaxis{\grfxaxis[dotted]{[]0.00;0.25;0.50;0.75;1.00[]}{[]\tiny$0.00$;\tiny$0.11$;\tiny$0.23$;\tiny$0.34$;\tiny$0.45$[]}}%
\def\setxgrid{\grfxgrid[dotted,gray]{[]0.00;0.25;0.50;0.75;1.00[]}{[]0.00;0.25;0.50;0.75;1.00[]}}%
\def\setyaxis{\grfyaxis[dotted]{[]0.00;0.25;0.50;0.75;1.00[]}{[]\tiny$2.14$;\tiny$2.32$;\tiny$2.50$;\tiny$2.69$;\tiny$2.87$[]}}%
\def\setlxaxis{\grfxaxis[dotted]{[]0.00;0.25;0.50;0.75;1.00[]}{[]\tiny$2.14$;\tiny$2.32$;\tiny$2.50$;\tiny$2.69$;\tiny$2.87$[]}}%
\def\setygrid{\grfygrid[dotted,gray]{[]0.00;0.25;0.50;0.75;1.00[]}{[]0.00;0.25;0.50;0.75;1.00[]}}%
\def\yexp{3}%
\def\settvyaxis{\grfyaxis[dotted]{[]0.00;0.25;0.50;0.75;1.00[]}{[]\tiny$2.56$;\tiny$3.92$;\tiny$5.28$;\tiny$6.65$;\tiny$8.01$[]}}%
\def\tvexp{4}%
\def\setmseyaxis{\grfyaxis[dotted]{[]0.00;0.25;0.50;0.75;1.00[]}{[]\tiny$14242.09$;\tiny$15976.44$;\tiny$17710.80$;\tiny$19445.15$;\tiny$21179.50$[]}}%

%% file: spect_sub_data.tex
\def\subrect{\draw[grfaxisstyle](-0.6562,0.7812) rectangle (0.0938,0.0312);}%
\def\isize{64}%
\def\jsize{64}%
\def\thetasize{60}%
\def\tsize{64}%